\documentstyle[aps,amssymb,amsmath,epsfig]{revtex}

\newcommand{\sK}{{\cal K}}

\newcommand{\sP}{{\cal P}}

\newcommand{\sL}{{\cal L}}
\newcommand{\sI}{{\cal I}}
\newcommand{\sB}{{\cal B}}
\newcommand{\sS}{{\cal S}}
\newcommand{\sV}{{\cal V}}
\newcommand{\sN}{{\cal N}}
\newcommand{\sD}{{\cal D}}

\begin{document} 

\title{\bf A spin foam model without bubble divergences}

\author{Alejandro Perez and Carlo Rovelli \\ {\it Centre de 
Physique
Th\'eorique - CNRS, Case  907, Luminy,
             F-13288 Marseille, France}, and \\
{\it Physics Department, University of Pittsburgh, 
             Pittsburgh, Pa 15260, USA}}

\maketitle

\begin{abstract}
We present a spin foam model in which the fundamental ``bubble
amplitudes'' (the analog of the one-loop corrections in quantum field
theory) are finite as the cutoff is removed.  The model is a natural
variant of the field theoretical formulation of the Barrett-Crane
model.  As the last, the model is a quantum BF theory plus an
implementation of the constraint that reduces BF theory to general
relativity.  We prove that the fundamental bubble amplitudes are
finite by constructing an upper bound, using certain inequalities
satisfied by the Wigner $(3n)j$-symbols, which we derive in the paper. 
Finally, we present arguments in support of the conjecture that the
bubble diagrams of the model are finite at all orders.

\end{abstract}

\newcommand{\SC}{\scriptscriptstyle} 
\newcommand{\SS}{\scriptscriptstyle}

\section{Introduction} 
To unlock the puzzle of quantum gravity, and understand quantum
spacetime, we have to learn how to treat general relativistic quantum
field theories in a  background independent 
fashion\cite{india}.  A surprising number of research directions aimed
at exploring background independent quantum field theory have
recently been converging towards the spin foam formalism
\cite{Reisenberger,Iwasaki,Baez,rr,Roberto,BarrettCrane}.

A spin foam model can be seen as a rigorous implementation of the
Wheeler-Misner-Hawking \cite{misner,haw} sum over geometries
formulation of quantum gravity.  The foam-like geometries summed over
are spin foams, or colored 2-complexes.  A 2-complex $J$ is a
(combinatorial) set of elements called ``vertices'' $v$, ``edges'' $e$
and ``faces'' $f$, and a boundary relation among these, such that an
edge is bounded by two vertices, and a face is bounded by a cyclic
sequence of contiguous edges (edges sharing a vertex).  A spin foam is
a 2-complex plus a ``coloring" $N$, that is an assignment of an
irreducible representation $N_{f}$ of a given group $G$ to each face
$f$ and of an intertwiner $i_{e}$ to each edge $e$.  The model is
defined by the partition function:
\begin{equation}
Z = \sum_{J}\ \sN(J) \sum_{N}\ \prod_{f\in J} \Delta_{N_{f}} \ \prod_{v\in
J} A_v(N),  \label{Z}
\end{equation}
where $\Delta_N$ is the dimension of the representation $N$, $A_v(N)$ is an
amplitude associated to vertices: a given function of the colorings of the
faces and edges adjacent to the vertex; and $\sN(J)$ is a normalization factor
for each 2-complex.

Let us mention a few of the research directions that, following very
different paths, have converged to models of this kind.  One of the
oldest models of this type is the Ponzano and Regge formulation of 2+1
gravity \cite{PonzanoRegge}.  The 2-complexes in this case are the
2-skeleta of the dual of Regge triangulations, and the vertex
amplitude turns out to be simply a Wigner $SU(2)$ 6j symbol.  In loop
quantum gravity \cite{loops}, spin foams emerge as histories (in
coordinate time) of quantum states of the geometry \cite{rr}, that
is, histories of spin networks\cite{spinnet,fs}.  In this case, the
vertex amplitude $A_v(c)$ is given by the matrix elements of the
Hamiltonian constraint.  In covariant lattice approaches, the sum over
colors corresponds to the integration over group elements associated
to links, expressed in a (``Fourier'') mode expansion over the group. 
In this case the vertex amplitude $A_v(c)$ is a discretized version of
(the exponential of) the Lagrangian density \cite{Reisenberg97,FK}. 
In topological field theories, the vertex function is a natural object
in the representation theory of the group $G$, satisfying a set of
identities that assure triangulation independence
\cite{t,Ooguri:1992b,CraneYetter,CraneYetter1}.  Finally, in the modifications of
topological quantum field theories related to quantum general
relativity \cite{BarrettCrane,Reisenberg97,iwa0}, the topological
field theory vertex amplitude is altered in order to incorporate a
quantum version of the constraints that reduces the BF topological
field theory \cite{BGGR,tqft-qg} to general relativity.

Spin foams are very much analogous to Feynman diagrams.  In references
\cite{dfkr,cm}, indeed, it is shown that the sum (\ref{Z}) can be
obtained as a Feynman expansion of a field theory over a group
manifold, whose interaction terms determine the vertex amplitude (see also
\cite{Roberto1}).  In this context, spacetime emerges from a Feynman expansion,
as in the old 2d quantum gravity matrix models, or zero-dimensional string
theory \cite{2d}.

As in the Feynman expansions of a conventional field theory, one
expects divergences to appear in the sum.  There are two types of
potential divergences: the ones associated to the sum over the colors
of a fixed 2-complex, and the ones associated with the sum over
2-complexes.  Here we consider only divergences of the first kind.  In
conventional Feynman diagrammatics, divergences are originated by
integrating over the momenta circulating along closed loops, because
momentum conservation at the vertices relates the momenta of adjacent
propagators.  In a spin foams model, the sum is over representations
associated to faces, and a constraint analogous to momentum
conservation is provided by the requirement of the existence of non
trivial intertwiners on the edges.  Consequently, divergences are
associated not to loops, as in quantum field theory, but rather to
{\em bubbles\/}: collections of faces forming a closed surface
\cite{coming}.  A way to control bubble divergences is to replace the
group $G$ with a $q$-deformed group, choosing $q$ such that $q^n=1$,
and to sum only over the finite set of proper representations of the
quantum group.  This is done, for instance, in the Turaev-Viro
\cite{turaev-viro} finite version of the Ponzano-Regge theory.  The
parameter $q$ plays the role of a cutoff, and the physical theory is
recovered by appropriately taking the $q\to 1$ limit.

The significance of the bubble divergences depends on whether the
model is topological.  A topological field theory is a diffeomorphism
invariant theory which does not have local degrees of freedom, but
only global ones.  General relativity in 3 spacetime dimensions and BF
theory are topological theories.  On the other hand, a non-topological
diffeomorphism-invariant theory is a theory, such as general
relativity in 4d, which is generally covariant but, nevertheless, has
local degrees of freedom (waves).  In the context of spin foam models,
the fact that a theory is topological is reflected in the fact that
the amplitude of a fixed 2-complex in (\ref{Z}) is 
independent from the 2-complex (triangulation independence ones the
manifold, i.e., the topology is fixed). More precisely, in general, in a
topological model, bubble amplitudes diverges in the $q\to 1$ limit, but
topological invariance implies that the sum over colorings is the same as the
one of a 2-complex in which the bubble has been removed, up to a divergent
overall factor that depends only on the cutoff.  In this case, therefore,
divergent diagrams do not provide any additional information.  The fact that a
topological theory does not have local degrees of freedom is reflected
in the triviality of all its ``radiative corrections" (bubble
diagrams).  The consequence is that the sum over 2-complexes is
trivial, and can be dropped, thus dropping all 2-complexes with
bubbles.

The situation is different in the non-topological context, and in
particular, for quantum general relativity.  In this case, the
``radiative corrections", that is, the bubble amplitudes, are the ones
that carry the information about the quantum behavior of the local
degrees of freedom of the theory.  In this context the sum over
2-complexes is necessary in order to capture all the degrees of freedom
of the theory, and bubble amplitudes are physically important.

Examples of non-topological models are the covariant expansion in
coordinate time obtained from loop quantum gravity, the Iwasaki\cite{iwa0}
and the Reisenberger\cite{BGGR} models, and the Barrett-Crane model.  The
Barrett-Crane model is a non-topological modification of a topological model:
4d $SO(4)$ BF theory, or the TOCY (Turaev-Ooguri-Barrett-Yetter) model
\cite{t,Ooguri:1992b,CraneYetter,CraneYetter1}. It is well known that 4d
$SO(4)$ BF theory has a peculiar relation with 4d Euclidean general relativity:
general relativity can be seen as a 4d $SO(4)$ BF theory plus a constraint,
which has an intriguing geometrical interpretation.  The Barrett-Crane model
is a modification of the TOCY model in which this constraint is implemented in
the quantum theory \cite{the quantum tetrahedrum,Barbieri}.  As a consequence
of the constraints, topological invariance is lost and the model acquires local
degrees of freedom.  Radiative corrections associated to bubble diagrams carry
non trivial physical information, but, unfortunately, diverge \cite{coming}. 
In order to be able to extract physical information from this model it is
necessary to deal with these divergences.

In a companion paper \cite{coming}, we have begun a general study of
the bubble amplitudes and the possible ways of renormalizing away
their infinities (see also \cite{Fotini}).  In particular, we have studied the
version of the Barrett-Crane model developed in \cite{dfkr}, in which the sum
over complexes is explicitly implemented by obtaining the spin foam model
from a field theory over a group manifold.  In the course of this
analysis, we have stumbled across a remarkable simple modification of
the action of the Barrett-Crane model which leads to finite
fundamental bubble amplitudes.\footnote{The precise meaning of the {\em
fundamental bubble diagrams} is explained below.  They represent the
simplest potentially divergent graphs in analogy to the one loop
corrections in standard QFT.} We present this model in the present
paper.  We also argue that all ``radiative corrections" in the model
are likely to be finite.  The model we present shares with the
original Barrett-Crane model the feature that makes it an intriguing
candidate for a quantum theory of (Euclidean) 4d general relativity:
that is, it is another implementation of the constraint that reduces
BF to GR. As a result, we obtain a theory, formally related to
Euclidean quantum general relativity, presumably finite at all orders
in the expansion over 2-complexes.

The paper is organized as follows. In section II, we recall the
field theory formulation of spin foam models by discussing the TOCY
model\cite{t,Ooguri:1992b,CraneYetter,CraneYetter1}, and the Barrett-Crane model; then we
introduce the new model. In section III, we study the model
in configuration space.  We define and compute the amplitudes
corresponding to the fundamental bubble diagrams.  We prove finiteness
of the 1-bubble diagram. In section IV, we study the model in momentum
space where we show that the interaction vertex of the theory is
the Barrett-Crane vertex.  We also give another proof of the finiteness of the
5-bubble amplitude.  In the appendix we review some known facts about the
representation theory of $SO(4)$, and we prove certain inequalities
satisfied by the Barrett-Crane vertex amplitude, the $6j$, and the
$15j$-symbols.  We use these inequalities to show the finiteness of
the fundamental bubble diagrams.  They generalize to the
$3Nj$-symbols.

\section{Definition of the model}

We begin by recalling the formulation of quantum BF theory, and of 
the Barrett-Crane Euclidean quantum gravity, as field theories on a 
group manifold. These were first developed by Ooguri in \cite{Ooguri:1992b}
and  by DePietri et al in \cite{dfkr}. Then we introduce the new model. 

\subsection{TOCY model as a QFT over a group manifold}

A spin foam model can be cast in the form of a field theory over a
group manifold \cite{dfkr,cm}.  The simplest of these theories is
given by the TOCY \cite{t,Ooguri:1992b,CraneYetter,CraneYetter1} topological model,
corresponding to the quantization of $SO(4)$ 4-dimensional BF theory. 
We begin by describing this model, which allows us to introduce
definitions relevant in the sequel.  The model is defined by the
action.  
\begin{eqnarray} 
    S[\phi] &=& \frac{1}{2} \int
 dg_1\ldots dg_{4} ~ \phi^{2}(g_1,g_2,g_3,g_4) \ +
\frac{\lambda}{5!} \int dg_1\ldots dg_{10} ~~\phi(g_1,g_2,g_3,g_4)\
\label{action} \\
\nonumber
&& \phi(g_4,g_5,g_6,g_7)\ \phi(g_7,g_3,g_8,g_9)\ 
\phi(g_9,g_6,g_2,g_{10}) \ \phi(g_{10},g_8,g_5,g_1).
\end{eqnarray} 
Here $g_{i}\in SO(4)$ and the field $\phi$ is a function over
$SO(4)^4$. All the integrals in this paper are in the normalized Haar measure. 
The field $\phi$ is required to be invariant under any permutation of its
arguments; that is,
$\phi(g_{1},g_{2},g_{3},g_{4})=\phi(g_{\sigma(1)},g_{\sigma(2)},
g_{\sigma(3)},g_{\sigma(4)})$, where $\sigma$ is any permutation of four
elements; and under the simultaneous right action of $SO(4)$ on its four
arguments: \begin{equation} 
    \label{ginv} 
    \phi(g_1,g_2,g_3,g_4) =
\phi(g_1g,g_2g,g_3g,g_4g) \ \ \ \ \forall g \in SO(4).  
\end{equation}

Let us introduce some simplification in the notation. 
We write $\phi(g_1,g_2,g_3,g_4)$ as $\phi(g_i)$, and we write the 
(\ref{action}) as 
\begin{equation}
    \label{actionsimple} 
    S[\phi]=\int dg_i \left[\phi(g_i)\right]^2 
    + {\lambda \over 5!} \int dg_i \left[\phi(g_i)\right]^5, 
\end{equation}
where, notice, the fifth power has to be interpreted as in
(\ref{action}).

Instead of requiring that the field satisfies property (\ref{ginv}),
we can also define the theory in terms of a generic field $\phi$ and
project it on the space of the fields satisfying (\ref{ginv}) by
integrating over the group.  That is, we can define 
the theory by the action
\begin{equation}
    \label{actionwithP} 
    S[\phi]=\int dg_i \left[ P_{g} \phi(g_i)
\right]^2 + {\lambda \over 5!} \int dg_i \left[ P_{g} \phi(g_i)
\right]^5, 
\end{equation} 
where the field is now arbitrary (except for the permutation 
symmetry), and the operator $P_g$ is defined by
\begin{equation} 
    \label{pg} P_{g}\phi(g_i)
\equiv \int d\gamma  \ {\phi}(g_i\gamma),
\end{equation}
where $\gamma \in SO(4)$.

Let us write the Feynman rules of the theory in coordinate space.  To 
this aim, we write the action as 
\begin{equation} 
    S[\phi] = \frac{1}{2} \int
dg_i \, d\tilde g_i \ 
\phi(g_i)\, \sK(g_i,\tilde g_i)\,  \phi(\tilde g_{i})
\ +
\frac{\lambda}{5!} \int 
dg_{ij}  \ 
\sV(g_{ij})
~\phi(g_{1j})~\phi(g_{2j})~\phi(g_{3j})~\phi(g_{4j})~\phi(g_{5j})
\label{actionexpanded} 
\end{equation} 
in the last integral, $i \neq j$.  Clearly, $\phi(g_{1j}) =
\phi(g_{12},g_{13},g_{14},g_{15})$, and so on.

\begin{figure}[h]
\centerline{{\psfig{figure=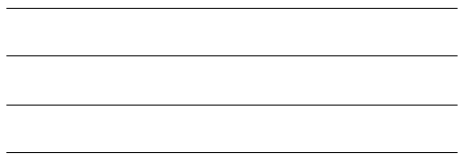,height=2cm}}}
\bigskip \caption{The structure of the propagator.}
\label{prop}
\end{figure}

The kinetic operator $\sK(g_i,\tilde g_i)$ can be written as
\begin{eqnarray} 
\label{propo1}
\sK(g_i,\tilde g_i)=\sum_{\sigma}\ \int d\gamma \
     \prod
\limits^4_{i=1}
\delta(g_i \gamma \tilde g^{-1}_{\sigma(i)}).
\end{eqnarray}
The propagator is the inverse of the kinetic operator, in the space of the
gauge invariant fields.  The operator $\sK$ is a projector (i.e.,
$\sK^2=\sK$): its inverse in the subspace of gauge invariant fields
corresponds to itself.  The propagator of the theory is then simply
\begin{eqnarray}\label{p}
\sP(g_i,\tilde g_i)= \sK(g_i,\tilde g_i). 
\label{propo2}
\end{eqnarray}
The propagator is formed by 4 delta functions (plus the symmetrization
and the integration over the group).  Its structure can therefore be
represented graphically as in Fig.\ \ref{prop}, in which every line
represents one of the delta functions in (\ref{p}). \\ 

\begin{figure}
\centerline{{\psfig{figure=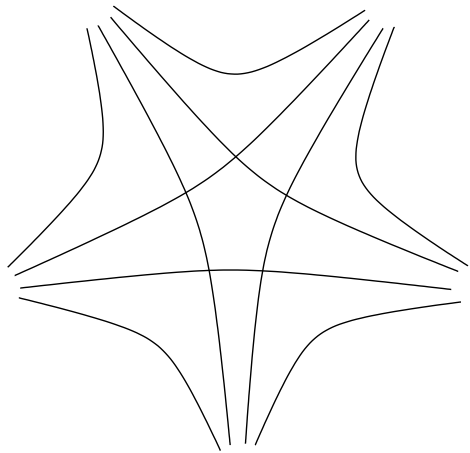,height=6cm}}}
\bigskip \caption{The structure of the interaction vertex.}
\label{vertex}
\end{figure}

The potential term can be written as 
\begin{equation} 
    \label{v5}
 \sV(g_{ij}) =  \frac{1}{5!}
\int d\beta_i \prod_{i < j} \delta(g_{ij}\beta^{-1}_i\beta_j 
g^{-1}_{ji}), 
\end{equation} where $\beta_i \in SO(4)$.
By introducing the notation $\rho_{ji}=g^{-1}_{ji}g^{}_{ij}$, which
satisfies the property $\rho^{-1}_{ij}=\rho_{ji}$, and rearranging the
arguments of the delta functions using the fact that
$\delta(g_1g_2)=\delta(g_2g_1)$, we can write the potential term as a
function of ten variables only
\begin{equation} 
    \label{v5due}
 \sV(\rho_{ij}) =  \frac{1}{5!}\int d\beta_i \prod_{i < j}
\delta(\beta_j\rho_{ji}\beta^{-1}_i). 
\end{equation}
The structure of the vertex is represented in Fig.\ \ref{vertex},
where each line represents one of the delta functions that appear in
the previous expression.\\
  
From now on we will use the Greek letter $\beta$ to denote the
symmetrization integration variables 
associated to vertices, while we reserve the letter $\gamma$ for
propagators.  

The Feynman diagrams of the theory are obtained by connecting the
five-valent vertices with propagators.  At the open ends of
propagators and vertices there are the four group variables corresponding to
the arguments of the field.  For a fixed permutation $\sigma$ in each
propagator, one can follow the sequence of delta functions with common
arguments across vertices and propagators.  On a closed graph, each
such sequence must close.  By associating a surface to each such sequence of
propagators, we construct a 2-complex \cite{dfkr}.  Thus,
by expanding in Feynman diagrams and in the sum over permutations in
(\ref{propo1}), we obtain a sum over 2-complexes.  In other words, a
2-complex is given by a certain vertices-propagators topology plus a
fixed choice of a permutation on each propagator. From now on, a 
``Feynman diagram", or simply a diagram, will denote one such 
2-complexes. 

According to the standard Feynman rules deduced from the form of the
action, the amplitude of any diagram will be given as an integral over
the internal coordinates -in this case $SO(4)$ group elements- of the
product of the corresponding propagators and vertices.  Thus, the
amplitude of any Feynman diagram is given by an integral of a product
of delta functions over a (compact) group manifold.  From (\ref{p})
and (\ref{v5}) we observe that the integrand defining this amplitude
correspond to a product of delta functions related to each other
according to Fig.\ \ref{prop} and \ref{vertex}.  

Working in configuration space, one can easily prove the topological
invariance of the theory using the Feynman rules derived from
(\ref{actionwithP})\cite{coming,Ooguri:1992b}. If one expands the action
(\ref{actionwithP}) in terms of irreducible representations using the
Peter-Weyl theorem, then the
amplitude of a Feynman diagram $J$ is
\begin{equation}
A(J)=\sum_N \prod_f \Delta_{N_f} \prod_v 15j(N_v),
\end{equation}
where $15j(N_v)$ denotes the $15j$-symbol
constructed out of the 10 colors of the surfaces meeting at the vertex
plus the five colors of the intertwiners corresponding to the five
 edges defining the vertex.  The sum goes over all possible compatible
colorings of faces and intertwiners in the 2-complex.

\subsection{The Barrett-Crane model in coordinate space}

Now we describe the Barrett-Crane model as a field theory over
$SO(4)^4$.  Consider the fundamental representation of $SO(4)$,
defined on $\Re^4$, and pick a fixed direction $\hat t$ in $\Re^4$. 
Let $H$ be the $SO(3)$ subgroup of $SO(4)$ that leaves $\hat t$
invariant.  We define the projector $P_{h}$  
\begin{equation} \label{ph}        
P_{h}\phi(g_i) \equiv \int dh_i \ {\phi}( g_ih_{i}),
\end{equation} where $h_i \in H$. From now on, the letter $h$ will always
indicate an element in $H=SO(3)$, while we use $g$, $\beta$, $\gamma$, and
$\rho$ for elements in $G=SO(4)$. The Barrett-Crane model is given by the
action:  \begin{equation}
    \label{bcm} 
    S[\phi]=\int
dg_i \left[ P_{g}P_{h} \phi(g_i) \right]^2 + {\lambda \over 5!} \int
dg_i \left[ P_{g}P_{h} \phi(g_i) \right]^5, 
\end{equation} 
where the notation is as in the previous section and the fields are 
assumed to be symmetric under permutations of their four arguments. 
In \cite{dfkr}, it is shown that the Feynman expansion of (\ref{bcm}) 
gives the Barrett-Crane spin foam model, summed over 2-complexes.  

Notice that the action (\ref{bcm}) is simply obtained by adding the
$P_{h}$ projection to the the action (\ref{actionwithP}) of the topological
TOCY. The projector $P_{h}$ projects the field over the linear
subspace of the fields that are constants on the orbits of $H$ in $G$, 
that is, that satisfy
\begin{equation}
    \label{gh} 
\phi(g_1,g_2,g_3,g_4)=\phi(g_1h_1,g_2h_2,g_3h_3,g_4h_4)
 \ \ \ \ \forall h_{i} \in H.  
\end{equation} 
When expanding the field in modes, that is, on the irreducible
representations of $SO(4)$, this projection is equivalent to
restricting the expansion to the representations in which there is a
vector invariant under the subgroup $H$ (because the projection
projects on such invariant vectors).  The representations in which such
invariant vectors exist are precisely the ``simple" representations,
namely the ones in which the spin of the self dual sector is equal to
the spin of the antiselfdual sector.  In turn, the simple
representations are the ones whose generators have equal selfdual and
antiself dual components, and this equality, under identification of
the $SO(4)$ generator with the $B$ field of $BF$ theory is precisely
the constraint that reduces $BF$ theory to GR. Alternatively, this constraint
allows one to identify the generators as bivectors
defining elementary surfaces in 4d, and thus to interpret the coloring
of a two-simplex as the choice of a (discretized) 4d geometry
\cite{Reisenberger,BarrettCrane,Barbieri}.

Using equations (\ref{pg}), and (\ref{ph}) it is straightforward to
compute the kinetic operator and the interaction vertex of the
Barrett-Crane model in coordinate space.  The kinetic operator of the
theory is 
\begin{equation}
    \label{pbc}
\sK(g_i,\tilde g_i)=\int  d\gamma dh_i \prod_{i}\delta(g_i h_i 
\gamma \tilde
h_i\tilde g^{-1}_i).  
\end{equation} 
The vertex is  
\begin{eqnarray}
\sV(g_{ij})&=& \frac{1}{5!}\int d\beta_i dh_{ij} 
\prod_{i < j}\  
\delta(g_{ji}^{-1}h_{ij}\beta^{-1}_i\beta_jh_{ji}g_{ij}).
\end{eqnarray}   
Notice that the kinetic operator $\sK$ is not a projector anymore 
(i.e., $\sK^2 \neq \sK$).  As a consequence, the propagator
$\sP=\sK^{-1}$ does not have a simple form in coordinate space.

If we expands the action (\ref{bcm}) in terms of irreducible
representations using the Peter-Weyl theorem, then the amplitude 
of a given Feynman diagram
$J$ is given by
\begin{equation}
A(J)=\sum_c \prod_f {\rm dim}(c_f) \prod_v \sB(c_v),
\end{equation}
where the sum is now only over simple representations of $SO(4)$, and
$\sB(c_v)$ denotes Barrett-Crane vertex amplitude.  This is given by
the $15j$-symbol constructed out of the 10 colors of the surfaces
meeting at the vertex using the Barrett-Crane intertwiners. See 
(\ref{bbcc}) and (\cite{dfkr}).

\subsection{The new model}

The idea at the basis of the new model is the same as in
Barrett-Crane: to modify the BF action (\ref{actionwithP}) by inserting the
projector (\ref{ph}) which implements the restriction of BF to GR.
However, this time we insert the projection in the interaction term
only, keeping the same kinetic term as in the BF theory. That is, we
define the new model by the action
\begin{equation} 
    \label{tope} 
    S[\phi]=\int
dg_i \left[ P_{g} \phi(g_i) \right]^2 + {\lambda \over 5!} \int dg_i
\left[ P_{g}P_{h}\phi(g_i) \right]^5 
\end{equation} 
or
\begin{equation} 
    \label{tope2} 
    S[\phi]=\int dg_i \left[ P_{g} \phi(g_i)
\right]^2 + {\lambda \over 5!} \int dg_i \left[ 
P_{g}P_{h}P_{g}\phi(g_i)
\right]^5. 
\end{equation} 
where $P_h$ and $P_g$ are defined in (\ref{pg}), and (\ref{ph})
respectively.  As for the Barrett-Crane model, if we drop $P_h$ from
the previous action we obtain the TOCY topological model of section
(II).  As we show below, the two forms of the action 
define the same theory, since the extra $P_g$ in the second expression
can be always absorbed into the $P_g$ of some propagator when
computing an amplitude.  The second form of the action 
simplifies the analysis of the theory in momentum space.  

From (\ref{tope}) the kinetic operator becomes 
\begin{eqnarray}\label{gaga}
\sK(g_i,\tilde g_i)=\int d\gamma dh_i \
 \prod_{i}  \delta(g_i \gamma \tilde
g^{-1}_i),
\end{eqnarray}
which corresponds to the projector into the space of gauge invariant
fields ($\sK^2=\sK$).  In this space its inverse is itself and the
propagator of the theory is simply $\sP=\sK$ as in the TOCY model. 
The vertex of the theory is 
\begin{eqnarray} 
    \label{vbc}
\sV(g_{ij})&=& \frac{1}{5!}\int d\beta_i d\hat\beta_i dh_{ij}\ 
\prod_{i < j} \delta(g^{-1}_{ji}\hat
\beta_i h_{ij}\beta^{-1}_i\beta_jh_{ji}\hat \beta^{-1}_jg_{ij}).
\end{eqnarray} 
The $\beta$ and $\hat \beta$
integration variables correspond to the two projectors $P_g$ in the
interaction.  In the following we will show that the $\hat \beta$
integration is redundant when computing any amplitude, and therefore
the two expressions in (\ref{tope}) and (\ref{tope2}) define the same theory. 
The key feature of this model is that, as we will argue in the following, it
does not contain divergences associated to the sum over colors for
fundamental bubble diagrams.

\section{Bubble amplitudes}

\subsection{Fundamental bubbles and Pachner moves}

As mentioned in the introduction, and discussed in detail in Ref.\
\cite{coming}, in a spin foam model divergences arise in  Feynman
diagrams containing bubbles, that is, closed surfaces.  Bubbles are
the spin foam analog of the loop diagrams of standard QFT.  To begin 
the analysis, we consider here only 2-complexes that are (the two 
skeleton of the) dual of regular triangulations.  Therefore, we will 
talk equivalently about triangulations or 2-complexes.

We define a {\em fundamental bubble diagram} as a bubble diagram obtained from
an elementary diagram without bubbles by means of the basic 4d Pachner moves
on the corresponding triangulation.  These are the simplest diagrams
presenting bubbles, and, in this sense, {\em fundamental bubble diagrams} are
analogous to one-loop diagrams in standard QFT, representing the basic
potentially divergent amplitudes in the model.

In four dimensions there are 3 possible Pachner moves --the 1-5, the
2-4, and the 3-3 Pachner moves respectively-- plus their inverses. 
Only the 1-5 and the 2-4 Pachner moves generate bubbles.
In terms of a 4d triangulation, the 1-5 move is defined by the split 
of a 4-simplex into five 4-simplices.  More precisely, we 
put a point $p$ in the interior of the 4-simplex with vertices $p_{i}, 
i=1,\ldots 5$, we add the five segments $(p,p_{i})$, the ten triangles 
$(p,p_{i},p_{j})$, the ten tetrahedra $(p,p_{i},p_{j},p_{k})$, and the 
five 4-simplices $(p,p_{i},p_{j},p_{k},p_{l})$ (where $i \ne j \ne k \ne l$). 

In the 2-4 move, we replace  the two 4-simplexes  $(a,p_1,p_2,p_3,p_4)$, and 
$(b,p_1,p_2,p_3,p_4)$, sharing the tetrahedron $(p_1,p_2,p_3,p_4)$ with the 
four 4-simplices $(a,b,p_{i},p_{j},p_{k})$ where $i\ne j\ne 
k=1,\ldots 4$. 

\begin{figure}
\centerline{{\psfig{figure=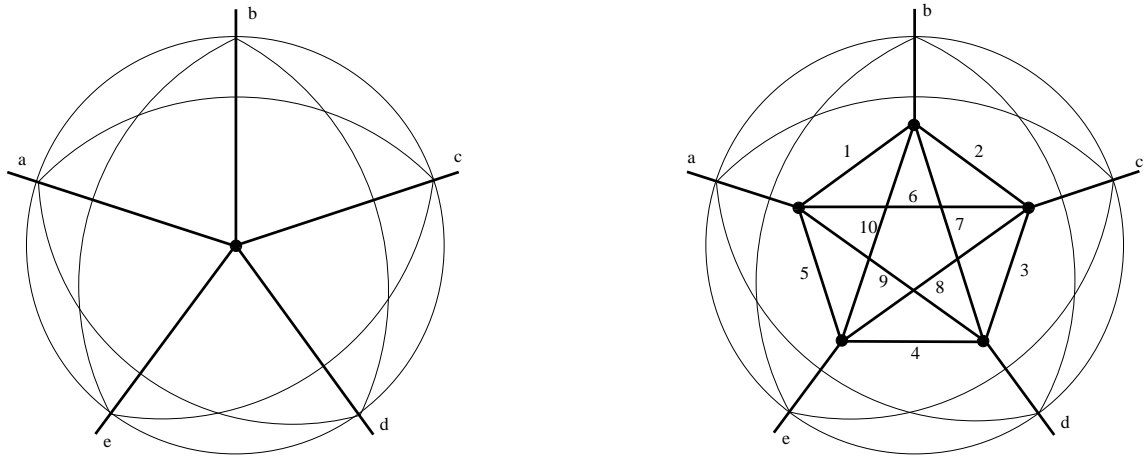,height=6cm}}} 
 \bigskip
\caption{The 1-5 Pachner move, on the right the fundamental 
5-bubble Feynman
diagram.} 
\label{15} 
\end{figure}

In terms of the 2-complex (the 2-skeleton of the dual of the
triangulation) which represent Feynman diagrams of our field theory,
this set of moves generate the two {\em fundamental bubble diagrams}. 
The 1-5 move in the dual picture is illustrated in Fig.\ \ref{15}.  We
denote the diagram on the right as the {\em 5-bubble diagram}.  The
vertices of the picture are dual to the 4-simplexes of the
triangulation; the edges of the picture are dual to the tetrahedra of
the triangulation; the surfaces (in fact, here, all the triangles) are
dual to the triangles of the triangulation.  

\begin{figure}
\centerline{{\psfig{figure=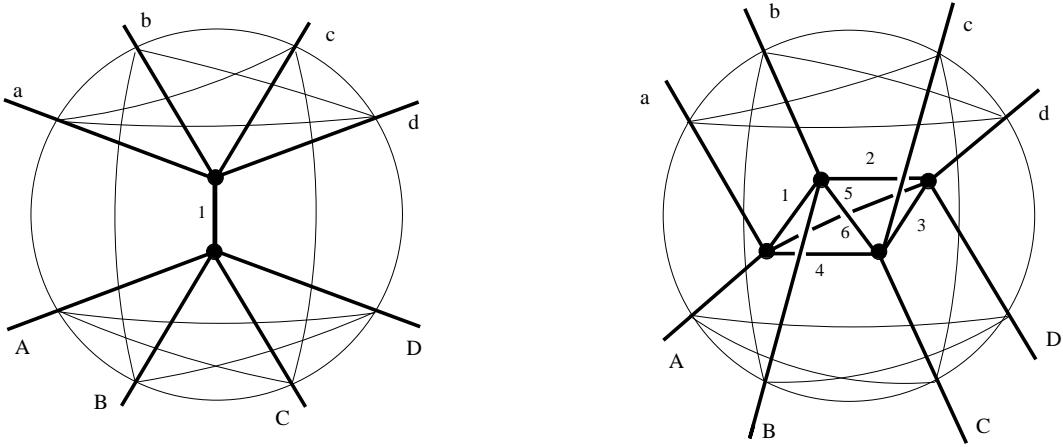,height=6cm}}}
\bigskip \caption{The 2-4 Pachner move, on the right the 
fundamental 1-bubble Feynman diagram.}
\label{24}
\end{figure}

The amplitude of a closed diagram is a number.  The amplitude of an
open diagram, that is, a diagram with a boundary, is a function of the
variables on the boundary, as for conventional QFT Feynman diagrams. 
The boundary of a 2-complex is given by a graph, where the nodes are
generated by the intersections of the edges with the boundary, and the
links are generated by the intersections of the surfaces with the
boundary.  To start with, the amplitude of the open diagram is a
function of 4 group arguments per each external line.  However,
consider a surface of an open 2-complex and the link $ab$ of the
boundary graph that bounds it.  Let $a$ and $b$ be the nodes on the
boundary graph that bound $ab$.  The surface determines a sequence of
delta functions that starts with one of the group elements in $a$, say
$g_{a}$ and ends with one of the group elements in $b$, say $g_{b}$. 
By integrating internal variables all these delta functions can be
contracted to a single delta function of the form $\delta(g_{a}\ldots
g_{b}^{-1})$.  This is of course a function of $g_{b}^{-1}g_{a}$.  We
can thus define the group variable $\rho_{ab}=g_{b}^{-1}g_{a}$,
naturally associated to the link $ab$, and conclude that the amplitude
of an open 2-complex is a function $A(\rho_{ab})$ of one group
element per each link of its boundary graph.  In ``momentum space",
the amplitude of the diagram is a function over the possible
colorings, in the sense of the spin networks, of the boundary graph. 
That is, if $s$ is a spin network given by a coloring of the boundary
graph,
\begin{equation}
A(s)=\int d\rho_{ab} \ \psi_{s}(\rho_{ab})\ A(\rho_{ab}), 
\end{equation}
where $\psi_{s}(\rho_{ab})$ is the spin network function\cite{spinnet}.
In Fig.\ \ref{15}, the thin lines in the picture represent the boundary graph
of the diagram, that is the intersection of the 2-complex with a 3-sphere
that bounds it.  This intersection is a graph on the
3-sphere.  Notice that the boundary remains the same after the move is
implemented.  We can think of the diagram on the right as a radiative
correction to the diagram on the left.  The analogous picture is shown
in Fig.\ \ref{24} for the 2-4 Pachner move.  We denote the diagram on
the right of Fig.\ \ref{24} as the {\em 1-bubble diagram}.\\


A straightforward calculation shows that the amplitudes of these
diagrams diverge in the cases of the TOCY and Barrett-Crane
model\cite{coming}.  In the following section, we compute the amplitudes of
the 5 and 1-bubbles.

\subsection{The 5-bubble amplitude}

We start by computing the amplitude $A_{5}$ corresponding to the
diagram on the right of Fig.\ \ref{15}.  We use the second form
of the action (\ref{tope2}) and show that the result is
independent of this choice. The amplitude is a function of the group
elements on the external propagators. It is obtained by writing the 
five vertices and the ten propagators in the diagram, and integrating 
on each propagator-vertex contraction. Notice that we are computing 
the amplitude of a fixed 2-complex, and therefore the permutation 
in (\ref{propo1}) is already fixed, and we do not have to sum over 
permutations.  

Let us introduce some notation.  We label the five vertices in Fig.\
\ref{15}  by means of an index $a$ that takes values from 1 to 5. 
The ten internal edges can then be labeled by unordered couples $(ab)$ of
indices.  For notational convenience we label as $(0a)$
the external edge emanating from the vertex $a$. 
Consider the $a$ vertex.  It is a function of twenty group
elements, four per leg, naturally paired by being in the same delta
function, or, equivalently, by relating to the same 2-complex
surface.  Consider the leg $(ab)$ of the vertex $a$.  We denote the 
group elements on this leg which is paired with the leg $(ac)$ as
$g_{(ab)(ac)}$. Notice that the first couple
of indices refer to the leg on which the group sits, while the second refers to
the leg  to which it is paired. The four group elements on external lines are 
denoted as $g_{(0a)(ab)}$, while the group  element on,
say, the $(ab)$ edge and paired to the external edge is  denoted
$g_{(ab)(a0)}$. Since there is one $h$ integration variable
corresponding to each of the configuration variables $g_{(ab)(bc)}$, we label
them according to the same rule. We denote the $\gamma$
integration variables in (\ref{gaga}) associated to propagators connecting the
vertex $a$ with the vertex $b$ as $\gamma_{ab}$ respectively. Given a vertex
there is one $\beta$ and one $\hat \beta$ per leg corresponding to the two
$P_g$'s in the interaction (see (\ref{tope2})). Therefore, the $\beta$
integration variables in (\ref{vbc}) are denoted as $\beta_{a(ab)}$, where the
first index denotes the vertex to which $\beta$ belongs, while the second
couple of indices denote the corresponding leg.   The amplitude  is then given
by  \begin{equation} \label{amplitude5}     A_{5}(g_{a(ab)}) = \int
[dg] \     \prod_{b}\ \sV(g_{(cb)(bd)})\     
\prod_{cd}\ \sP(g_{(cd)(ce)},g_{(cd)(de)}),     \end{equation} where $[dg]$
denotes the integration over internal group elements (namely, $g_{(ab)(bc)}$
with $a,b \ne 0$), while the indices $c,d,e$ in the vertices and propagators
can take the value $0$ as well.  As argued in the last section,   the
$g_{(0a)(ab)}$'s enter this expression only in the combination
$\rho_{ab}=g^{-1}_{(0a)(ab)}g^{}_{(0b)(ab)}$. 

We have to insert the 
values (\ref{vbc}) of the vertex amplitude and the value
(\ref{propo1},\ref{propo2}) of the propagator (without the sum over
permutations) in this expression. Explicitly, and with the correct index
structure we need, these are  \begin{equation}
\label{vertexin}
  \sV(g_{(cb)(bd)}) =
 \frac{1}{5!}\int [d\beta] [d\hat\beta] [dh] \!\!\!
\prod_{(cb) < (bd)} \!\!\!\! \delta(g^{-1}_{(bd)(cb)}\,\hat
\beta_{b(cb)} h_{(cb)(bd)}\beta^{-1}_{b(cb)}\beta_{b(bd)} 
h_{(bd)(cb)}\hat \beta^{-1}_{b(bd)} \,
g_{(cb)(bd)})
\end{equation}
and
\begin{eqnarray} 
    \sP(g_{(cd)(ce)},g_{(cd)(de)})=\int [d\gamma]
     \prod
\limits_{e=0,\ldots,5, e\ne c,e\ne d}
\delta(g_{(cd)(ce)} \gamma_{cd} \tilde g^{-1}_{(cd)(de)}),
\label{propoin}
\end{eqnarray}
where $[d\beta]$, $[d\hat \beta]$, $[dh]$, and $[d\gamma]$ denote integration
over all the $\beta_{a(ab)}$, $\hat \beta_{a(ab)}$, $h_{(ab)(bc)}$, and
$\gamma_{ab}$ variables respectively. We recall that all $h$'s are in $SO(3)$,
while all $\beta$'s, $\gamma$'s, $g$'s, and $\rho$'s are in $SO(4)$. Inserting
(\ref{vertexin}) and (\ref{propoin}) in (\ref{amplitude5}) we obtain a
complicated multiple integral of delta functions, which we now analyze by
breaking it into pieces.

The key observation is that the delta functions appear in sequences, 
corresponding to the boundaries of the surfaces of the faces of the 
2-complex, or to the connected lines obtained by replacing vertices 
and propagators in Fig.\ \ref{15} with the vertex and propagators given in
Fig.\ \ref{prop} and \ref{vertex}. It is easy to see that  there are two
kinds of such sequences, corresponding to the two kind of surfaces in 
the 2-complex. These are illustrated in Fig.\ 5. 

\begin{figure}[h]\label{a15}
\centerline{\hspace{0.6cm} 
\psfig{figure=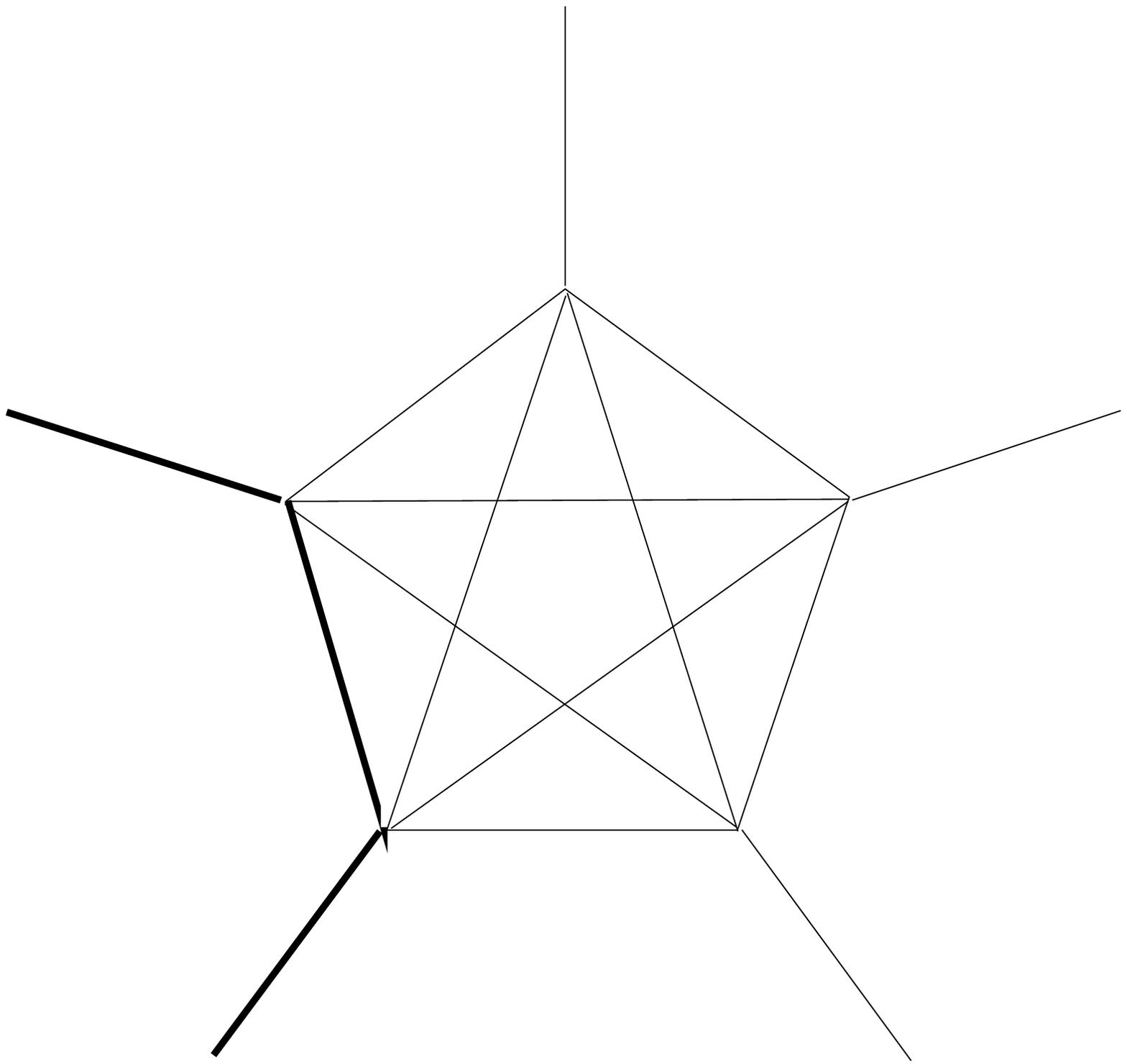,width=5cm,angle=0} 
\,\,\,\,\,\,\,\,\,\,\,\,\,\,\,\,\,\,\,\, 
\,\,\,\,\,\,\,\,\,\,\,\,\,\,
\psfig{figure=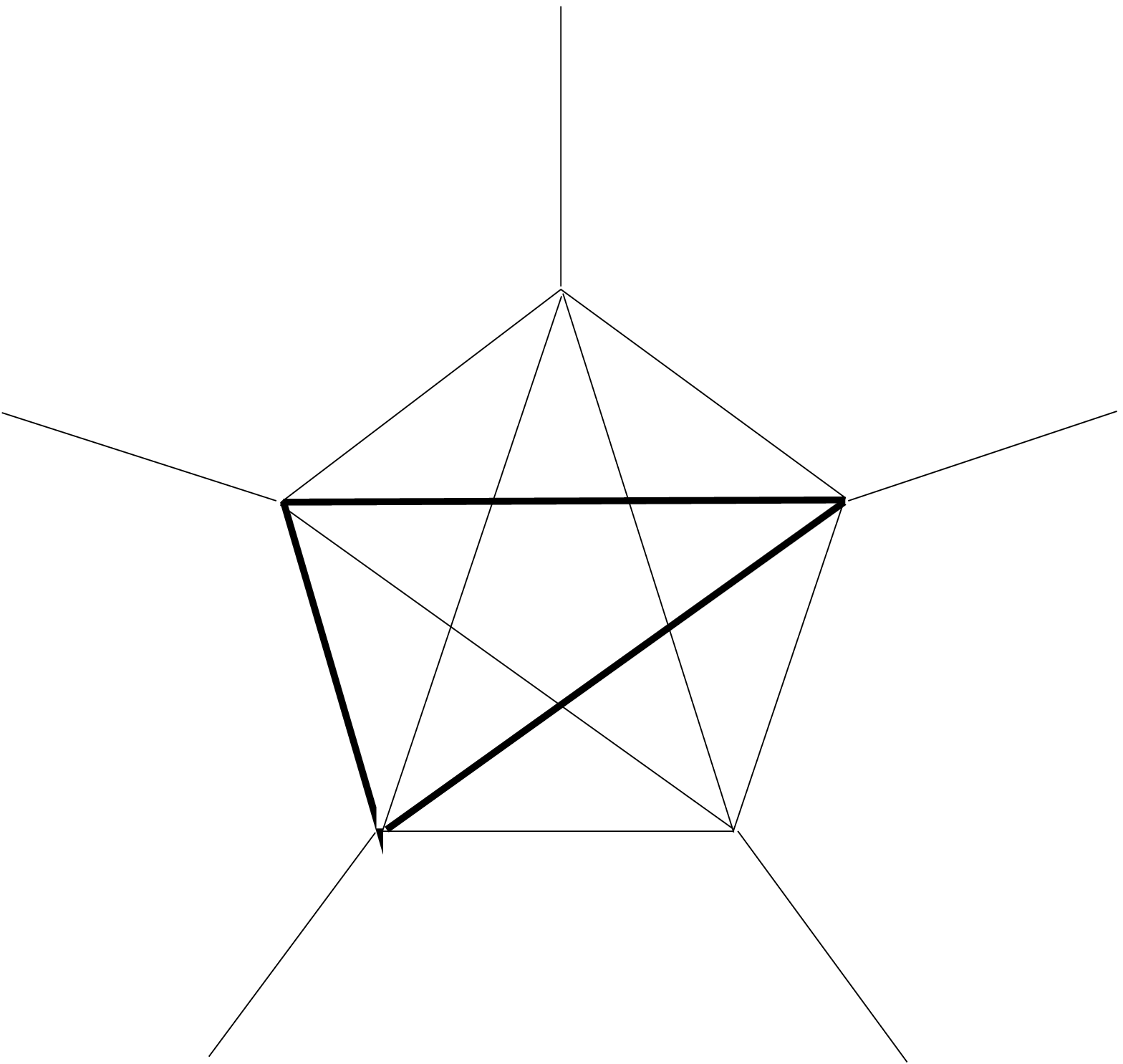,width=5cm,angle=0}}    
\begin{center}    
\parbox{15cm}{\caption{Two possible terms contributing to the
${\em A}_5$ amplitude}}     
\end{center}  
\end{figure}

The terms of the first kind follow the pattern shown in the
diagram on the left of Fig.\ 5; we denote them as wedges.  The
ten terms of the second kind have the form of the diagram shown on the
right, and we denote them as faces. There are ten terms of the first kind 
and ten terms of the second kind.  
Consider the term of the first kind. Let $(ab)$ be the link in this 
term. From the definitions of
the vertex and propagator respectively we have that the corresponding 
sequence of delta functions is   
\begin{eqnarray} 
\nonumber &&\delta(g_{(0a)(ab)}\hat\beta^{}_{a(0a)}h_{(0a)(ab)}
\beta^{-1}_{a(0a)}
\beta^{}_{a(ab)}h_{(ab)(0a)}
\hat\beta^{-1}_{a(ab)}g^{-1}_{(ab)(0a)})
\delta(g_{(ab)(0a)}\gamma_{(ab)}g^{-1}_{(ab)(0b)})\\
&&\ \ \ \ \ \ \ \ \ \ \ \ \ \ \ \ \ \ \ \ \ \ \ \ \ \ \ \ \ \ \ \ \ \ \ \ \ 
\ \ \ \ \ \ \ \ \ \ \ \ 
\delta(g_{(ab)(0b)}\hat\beta^{}_{b(ab)}h_{(0b)(ab)} \beta^{-1}_{b(ab)}
\beta^{}_{b(0b)}h_{(0b)(ab)}
\hat\beta^{-1}_{b(0b)}g^{-1}_{(b0)(ab)}),  
\end{eqnarray} 
where the first and the third delta functions come
from the vertex $a$ and the vertex $b$ respectively, and the delta in
the middle comes from one of the four deltas in the propagator $ab$
connecting the two vertices.  By integrating over the two variables
that concatenate the delta functions, namely $g_{(ab)(0a)}$ and
$g_{(ab)(0b)}$, we obtain the quantity
\begin{equation} 
E_{ab}\equiv\delta(\hat\beta^{}_{b(ab)}h_{(ab)(0b)}\beta^{-1}_{b(ab)}
\beta^{}_{b(0b)}h_{(0b)(ab)}\beta^{-1}_{b(0b)}\rho_{ba} 
\hat\beta^{}_{a(0a)}
h_{(0a)(ab)}\beta^{-1}_{a(0a)}\beta^{}_{a(ab)}h_{(ab)(0a)}
\hat\beta^{-1}_{a(ab)}\gamma^{}_{ab}),  
\label{Ebc} 
\end{equation} 
We have rearranged the
argument of the delta function using that
$\delta(g_1g_2)=\delta(g_2g_1)$, and we have defined $\rho_{ab} \equiv
g^{-1}_{(0a)(ab)} g_{(0b)(ab)}$.  

The second kind of terms, illustrated on the right of Fig.\ \ref{a15},
come from the ten internal faces in the 5-bubble diagram.  Consider 
the face bounded by the three vertices $abc$. The corresponding term 
gives the sequence of deltas
\begin{eqnarray}
\label{Fbc}
&&\nonumber \delta(g_{(ac)(ab)}
\hat\beta^{}_{a(ca)}h_{(ac)(ab)}\beta^{-1}_{a(ca)}
\beta^{}_{a(ab)}h_{(ab)(ac)}\hat\beta^{-1}_{a(ab)}g^{-1}_{(ab)(ac)})
\delta(g_{(ab)(ac)} \gamma_{(ab)} g^{-1}_{(ab)(bc)})\\
&& \nonumber 
\ \ \ \ \ \ \ \ \ \ \ \ \ 
\delta(g_{(ab)(bc)}\hat\beta^{}_{b(ab)}h_{(ab)(bc)}
\beta^{-1}_{b(ab)}\beta^{}_{b(bc)}h_{(bc)(ab)}
\hat\beta^{-1}_{b(bc)}g^{-1}_{(bc)(ab)})
\delta(g_{(bc)(ab)}\gamma^{}_{(bc)}g^{-1}_{(bc)(ac)})\\
&& \ \ \ \ \ \ \
\ \ \ \ \ \ \ \ \ \ \ \ \ \ \ \ \ \ 
\ \ \ \ \ \ \ \ \ \ \ \ \ \delta(g_{(bc)(ac)}\hat\beta^{}_{c(bc)}h_{(bc)(ac)}
\beta^{-1}_{c(bc)}\beta^{}_{(c)ca}h_{(ac)(bc)}
\hat\beta^{-1}_{c(ca)}g^{-1}_{(ac)(bc)}) \delta(g_{(ac)(bc)} \gamma^{}_{(ac)}
g^{-1}_{(ac)(ab)}) \end{eqnarray}
Again, we can integrate over the intermediate group elements,
$g_{(ac)(ab)}$, $g_{(ab)(ac)}$, $g_{(ab)(bc)}$, $g_{(bc)(ab)}$,
$g_{(bc)(ac)}$, and $g_{(ac)(bc)}$ obtaining
\begin{eqnarray}
&& \nonumber F_{abc}\equiv\delta(\underbrace{ \beta_{a(ab)}
h_{(ab)(ac)}\hat\beta^{-1}_{a(ab)}\gamma_{(ab)}
\hat\beta^{}_{b(ab)}h_{(ab)(bc)}
\beta^{-1}_{b(ab)}}_{\rho^{\SC abc}_{ab}}\beta_{b(bc)}
h_{(bc)(ab)}\hat \beta^{-1}_{b(bc)}\\
&&\ \ \ \ \ \ \ \ \ \ \
\ \ \ \ \ \ \ \ \ \ \ \ \ \ \ \ \ \ 
\ \ \ \ \ \ \ \ \ \ \ \ \ \gamma^{}_{(bc)}
\hat\beta^{}_{c(bc)}h_{(bc)(ac)}\beta^{-1}_{c(bc)}\beta_{c(ca)}
h_{(ac)(bc)}\hat\beta^{-1}_{c(ca)}\gamma^{}_{(ac)}\hat\beta_{a(ca)}
h_{(ac)(ab)}\beta^{-1}_{a(ca)}),
\end{eqnarray} 
Notice that the $\gamma_{ab}$'s associated to the internal propagators
appear surrounded by $\hat\beta$'s in the same way in $E_{ab}$ and in
$F_{abc}$; therefore we can reabsorb all the $\hat \beta$'s by
redefining the integration variables $\gamma_{ab}$ (that is
$\gamma_{ab} \rightarrow \hat \beta^{-1}_{(a)ab}\gamma_{ab}
\hat\beta_{(b)ab}$) using the right-left invariance of the Haar
measure.  This shows that the $P_g$ on the right in the interaction
term of the action (\ref{tope2}) is redundant and that the two forms
(\ref{tope}) and (\ref{tope2}) of the action are equivalent.

Using the definitions (\ref{Ebc}) and (\ref{Fbc}), we can rewrite the 
amplitude (\ref{amplitude5}) as 
\begin{equation}
    A_{5}(\rho_{ab})
   = \int [d\gamma]\ [d\beta] \ [d\hat \beta]\ [dh] \   
   \prod_{ab}\ E_{ab}
   \prod_{abc}\ F_{abc}.    
\end{equation}
We can eliminate the $E_{ab}$'s by integrating over the 
$\gamma_{ab}$'s. This gives
\begin{eqnarray}\label{gordo}
\gamma_{ab}=
\hat\beta_{a(ab)}h_{(ab)(0a)}\beta^{-1}_{a(ab)}\beta^{}_{a(0a)}h_{(0a)(ab)}
\hat\beta^{-1}_{a(0a)}\rho_{ab} 
\hat\beta_{b(0b)}h_{(0b)(ab)}\beta^{-1}_{b(0b)}
\beta_{b(ab)}h_{(ab)(0b)}\hat\beta^{-1}_{b(ab)},  
\end{eqnarray} 
We substitute these values into the remaining deltas; we define 
(see the emphasized term in equation (\ref{Fbc}))  
\begin{eqnarray}\label{rb}
\rho^{\SC abc}_{ab}=\beta^{}_{a(ab)}
h_{(ab)(ac)}h_{(ab)(0a)}\beta^{-1}_{a(ab)}
\beta^{}_{a(0a)}h_{(0a)(ab)}\hat\beta^{-1}_{a(0a)}
\rho_{ab}\hat\beta^{}_{b(0b)}h_{(0b)(ab)}\beta^{-1}_{b(0b)}\beta^{}_{b(ab)}
h_{(ab)(0b)}h_{(ab)(bc)}\beta^{-1}_{b(ab)}. 
\end{eqnarray} 
We can redefine the $h_{(ab)(bc)}$ variables, by absorbing the 
$h_{(ab)(0b)}$ into them.  We substitute these expression in the $F$'s, 
and the  amplitude becomes 
\begin{equation} 
    \label{rr} 
{A}_5(\rho_{ab})= \int [d\beta] \ [d\hat \beta]\ [dh] \
\prod_{\SC abc} \delta(\rho^{\SC abc}_{ab}\rho^{\SC abc}_{bc}\rho^{\SC
abc}_{ca}).  
\end{equation} 

In order to show that ${\em A}_5(\rho_{ab})$ is finite, it is
sufficient to show that the ``vacuum" value of the amplitude is
finite\cite{coming}.  In momentum space this is the value of
the amplitude when the colors of the boundary lines are zero.  In
configuration space this value is obtained by integrating the
amplitude (\ref{rr}) over $\rho_{ab}$ (Integration over the external
group variables projects the amplitude to the trivial
representation).  We want thus to show that 
\begin{equation} 
    \label{gordita} 
    \sI_5 = \int [d\rho] \ {A}_5(\rho_{ab}) < \infty. 
\end{equation} 
The amplitude when the external colors are not all zero is given by
the integral of ${A}^{(bc)}_5(\rho)$ times the appropriate
spin-network state, which is a bounded function over the group.  The
result of this integration is finite if (\ref{gordita}) holds.

If we insert (\ref{rb}) and (\ref{rr}) into 
(\ref{gordita}), we see that several of the variables can 
be absorbed into redefinitions of integration variables. 
In particular, we can redefine  $\rho_{ab}
\rightarrow \beta^{}_{a(0a)}h_{(0a)(ab)}\hat\beta^{-1}_{a(0a)}
\rho_{ab}\hat\beta^{}_{b(0b)}h_{(0b)(ab)}\beta^{-1}_{b(0b)}$ in 
(\ref{gordo}), and obtain the following simpler expression
\begin{eqnarray}\label{gorditabis}
   &&\nonumber \sI_5 = \int [d\rho]
\ [d\beta] \ [dh]  \ 
\prod_{abc}\delta( \beta^{}_{a(ab)}
h_{(ab)(bc)}\beta^{-1}_{a(ab)}
\rho_{ab}\beta^{}_{b(ab)}h_{(cb)(ab)}\beta^{-1}_{b(ab)}\\
&&\ \ \ \ \ \ \ \ \ \ \ \ \ \ \ \ \ \ \ \ \ \  \ \beta_{b(bc)}
h_{(ab)(bc)}\beta^{-1}_{b(bc)}
\rho_{bc}\beta^{}_{c(bc)}h_{(bc)(ab)}\beta^{-1}_{c(bc)}
\beta_{c(ca)}
h_{(ca)(cb)}\beta^{-1}_{c(ca)}
\rho_{ca}\beta^{}_{a(ca)}h_{(cb)(ca)}\beta^{-1}_{a(ca)}). 
\end{eqnarray} 
We can write this as 
\begin{equation}
   \sI_5 = \int [d\rho]
\ [d\beta] \ [dh] \ 
\prod_{abc}\delta(\tilde\rho^{\SC abc}_{ab} 
\tilde\rho^{\SC abc}_{bc} 
\tilde\rho^{\SC abc}_{cd}), 
\end{equation}
where 
\begin{equation}
\label{tilderho}
    \tilde\rho^{\SC abc}_{ab}=
\beta^{}_{(a)ab}
h_{(ab)(bc)}\beta^{-1}_{(a)ab}
\rho_{ab}\beta_{(b)ab}
h_{(bc)(ab)}\beta^{-1}_{(b)ab}. 
\end{equation} 

Now, in order to simplify this integral, we observe that the
integrand in (\ref{gorditabis}) is gauge invariant under the gauge
transformation $\rho_{ab} \rightarrow \mu^{}_a \rho_{ab}\mu^{-1}_b$
with $\mu_a \in SO(4)$, since the $\mu_a$'s can be absorbed by
redefining the $\beta$'s.  The integral is
therefore equal to the gauge fixed integral at a particular gauge,
times the volume of the gauge orbit (which is unity since we are working with
the normalized Haar measure).  We fix the gauge by requiring
$\rho_{12}=\rho_{13}=\rho_{14}=\rho_{15}=1$.  Taking this into
account, we can integrate away the variables $\rho_{\hat b\hat
c}$, where an index with hat takes the values 2 to 5 only. The delta 
functions -associated to faces in which 2 of the 3 $\rho$'s have been gauge
fixed- fix the value of these variables to  \begin{eqnarray} 
&& \nonumber \rho_{\hat b\hat c}=
\beta_{\hat b(\hat b\hat c)}
h_{(\hat b\hat c)(\hat c 1)}\beta^{-1}_{\hat b(\hat b
\hat c)}\beta^{}_{\hat b(1\hat b)} h_{(\hat b \hat c)(1 \hat
b)}\beta^{-1}_{\hat b(1\hat b)} \beta_{1(1\hat b)} \hat h_{(1\hat b)(\hat
b\hat c)} \beta^{-1}_{1(1\hat b)}\\
&& \ \ \ \ \ \ \ \ \ \ \ \ \ \ \ \ \ \ \ \ \ \ \ \ \ \
 \ \ \ \ \ \ \ \ \ \ \ \ \beta_{1(\hat c1)} 
h_{(\hat c 1)(1\hat b)}
\beta^{-1}_{1(\hat c1)}
\beta_{\hat c(\hat c1)}
h_{(1\hat b)(\hat c1)}\beta^{-1}_{\hat c(\hat c1)}
\beta_{\hat c(\hat b\hat c)}
h_{(\hat c 1)(\hat b\hat c)}\beta^{-1}_{\hat c(\hat b\hat c)}. 
\end{eqnarray} 
Inserting this in (\ref{tilderho}), the external
$\beta's$ simplify and two $h$' can be collapsed 
into one by redefinition.  
There remain only the four delta functions associated
to the faces $\hat a\hat b\hat c$. Each 
$h$ variable turns out to be sandwiched between $\beta$ and 
$\beta^{-1}$. Thus, integrating in such an $h$ variable is indeed 
integrating over the subgroup $\beta H$ of $SO(4)$, formed by the 
elements that leave $\beta \hat t$ invariant. Taking all this into 
account, we conclude
\begin{eqnarray} 
     \sI_5
&=& \int d\beta_{b(cd)}\ \int_{\beta_{(ab)(cd)}H} d{\frak h}_{(ab)(cd)} \
\prod_{abc=2,\ldots,5} 
\delta(\tilde \rho^{\SC abc }_{ab}\tilde \rho^{\SC abc }_{bc}\tilde \rho^{\SC
abc }_{cd}), \label{integral}
\end{eqnarray}
with 
\begin{equation}
\label{tilderho2}
\tilde\rho^{\SC abc}_{ab}= {\frak h}_{(bc)(cd)} {\frak h}_{(ab)(bc)}
{\frak h}_{(ab)(ac)} {\frak h}_{(ca)(ab)} {\frak h}_{(ac)(bd)} {\frak
h}_{(bc)(bd)},\end{equation} 
where ${\frak h}_{(ab)(bc)} \in \beta_{(ab)(bc)} H$
(each ${\frak h}$ corresponds to a different $SO(3)$ subgroup).

As a check, imagine now that we are dealing with BF
theory.  To obtain BF theory, we have simply to drop the integrations over
$h$'s.  The resulting divergence is immediately $(\delta(1))^{4}$, which is the
correct divergent factor for the 1-5 pachner move in BF theory
\cite{Ooguri:1992b,coming}.  Does the integrations over the subgroup
$H$ absorb all the divergences?  If it wasn't for the integration of
the $\beta$'s, the answer would be no, because we would simply obtain
a divergence proportional to the fourth power of the value on the
origin of the delta on $G/H$.  However, the combined integration over
the $h$'s and the $\beta$'s is sufficient to absorb all the
divergences.

In fact, the integral we are studying is over a compact domain.  Therefore
divergences can only come from particular points where the integrand
diverges.  The point where the most serious divergences can appear is
the origin $\beta_{(ab)(cd)}=h_{(ab)(cd)}=1$.  Let us study the
behavior of the integral around this point.  To this aim, it is
sufficiently to study an arbitrarily small neighborhood of this
point.  On an arbitrarily small neighborhood of the origin, we can
describe the group by means of its algebra.  We can thus replace the
group integral with an integral over the algebra.  To the order we are
interested, we can take the Lebesgue measure over the algebra, since
the Jacobian between this measure and the measure induced by the Haar
measure on the group goes smoothly to 1 on the origin.  Group elements
can be written as exponents of algebra elements, and products of group
elements can be expanded to first order around the origin, giving
commutators in the algebra.  The four $SO(4)$ delta functions give 24
1-dimensional deltas.  It is lengthy (because of the many variables),
but completely straightforward, to see that the 24 resulting
1-dimensional delta functions in the previous expression are not
redundant and therefore the value of the amplitude is finite. We will 
present the details of the calculation elsewhere.  

A simple proof of the finiteness of the previous amplitude will be
given in terms of the mode expansion in momentum space.

\subsection{The 1-bubble amplitude}

We now consider the amplitude of the diagram 
on the right of Fig.\ \ref{24}. This is another divergent
amplitude in BF theory and in the Barrett-Crane model. We 
now show that the amplitude is finite in the new model. 
The pattern for analyzing this amplitude is the same as in the previous case,
but simpler. We only sketch here the key steps of 
the calculation, leaving the details to the reader. 
Integration over the six upper external wedges (see Fig.\ \ref{24}) implies, in 
analogy to
the previous case, that
\begin{eqnarray}
\gamma^{}_{ab}=\hat\beta^{}_{a(ab)}h_{(ab)(0a)}
\beta^{-1}_{a(ab)}\beta^{}_{a(0a)}h_{(0a)(ab)}
\hat\beta^{-1}_{a(0a)}\rho_{ab} \hat
\beta^{}_{b(0b)}h_{(0b)(ab)}\beta^{-1}_{b(0b)}\beta^{}_{b(ab)}
h_{(ab)(0b)}\hat\beta^{-1}_{b(ab)}. 
\end{eqnarray}
substituting this into the remaining deltas we obtain
(schematically) 
\begin{eqnarray}
\nonumber {\em A}_4(\rho)=  \int [d\beta] [dh]
\prod_{ab{\SC A}{\SC B}}\!\!& \delta &(\beta^{}_{a(ab)}
h_{({\SC B}{\SC 
A})(0 {\SC
A})}\beta^{-1}_{a(ab)}\beta^{}_{a(0a)}h_{(0a)(ab)}\rho_{ab}h_{(0b)(ab)}
\beta^{-1}_{b(0b)}
\\ \nonumber  
&& \beta^{}_{b(ab)}\hat h_{({\SC B}{\SC
A})(0{\SC B})}\beta^{-1}_{b(ab)}\beta_{b({0\SC B})}h_{(0{\SC B})({\SC B}{\SC 
A})}\rho_{{\SC B}{\SC A}}h_{(0{\SC A})({\SC B}{\SC A})}\beta^{-1}_{a(0{\SC
A})})\\  \prod_{abc}
\!\!& \delta&(\rho^{\SC (abc)}_{ab}\rho^{\SC
(abc)}_{bc}\rho^{\SC (abc)}_{ca}) \prod_{a{\SC A}}
\delta(\beta^{}_{a(0a)}h_{(0a)(0{\SC A})}\rho_{a{\SC A}}
h_{(0{\SC A})(0a)}\beta^{-1}_{a(0{\SC A})}), \end{eqnarray} Now we want to 
show that the
``vacuum bubble'' is finite, namely: \begin{equation} 
\sI_{4}=\int [d\rho]\  {\em
A}_4(\rho) < \infty. \end{equation}
Integrating over 
$\rho_{AB}$
and $\rho_{aA}$ this reduces to 
\begin{equation}\sI_{4} = \int 
[d\rho] [d\beta] [dh]
\prod_{abc} \delta(\rho^{\SC (abc)}_{ab}\rho^{\SC 
(abc)}_{bc}\rho^{\SC
(abc)}_{ca}).
\end{equation} 
As in the previous case we can fix the
gauge by means of the conditions 
$\rho_{12}=\rho_{13}=\rho_{14}=1$. We then
integrate over the remaining $\rho$'s which 
eliminate three 
of the four delta functions, and 
fix the
value of the remaining $\rho_{cd}$ (for $c,d \neq 1$) to 
\begin{eqnarray} &&\nonumber \rho_{bc}=\beta^{}_{b(bc)}
h_{(bc)(1b)}\beta^{-1}_{b(bc)}\beta^{}_{b(ab)} 
h_{(ab)(ac)}\beta^{-1}_{b(ab)}\beta^{}_{a(ab)} 
h_{(ab)(bc)}\beta^{-1}_{a(ab)}\\
&&\ \ \ \ \ \ \ \ \ \ \ \ \ \ \ \ \ \ \ \ \ \ \ \ \ \ \ \ \ 
\ \ \ \ \ \ \ \ \ \ \ \ \ \ \ \ \ \ \ 
\beta^{}_{a(ca)} 
h_{(ca)(ab)}\beta^{-1}_{a(ca)}\beta^{}_{c(ca)}
h_{(ca)(cb)}\beta^{-1}_{c(ca)}\beta^{}_{c(bc)}
h_{(bc)(ca)}\beta^{-1}_{c(bc)}. 
\end{eqnarray}
Finally, $\sI_{4}$ reduces to an integral of single delta function. 
This integral is the integration of the distribution $\sD(g_1,g_2,g_3)$
defined in (\ref{ii}).  By lemma(1) in the appendix, the integral is
finite.

\section{Spin foam formulation}

Using Peter-Weyl theorem, we can analyze the new model in momentum
space.  This analysis explicitly connects our results to the spin foam
formalism.  In particular, we show that the vertex amplitude of the
theory is essentially the same as the one in the Barrett-Crane model, up
to a rescaling, that can be viewed as an edge amplitude.  We also
present here another proof that the amplitudes calculated in the
previous section are finite, and finally we present arguments 
supporting the conjecture that amplitudes are finite at any order. 

According to the Peter-Weyl theorem, given an $\sL^2[G]$ (Haar-square 
integrable) function $\phi(g)$ over the group $G=SO(4)$,  we
can expand it in terms of the  matrices
$D^{(\Lambda)}_{\alpha\beta}(g)$ of the irreducible representations 
$\Lambda$ of $SO(4)$; that is,  
\begin{equation} 
    \phi(g) = \sum_\Lambda \
\phi^{\Lambda}_{\alpha\beta} \ D^{(\Lambda)}_{\alpha\beta}(g). 
\end{equation}
We begin by analyzing the kinetic term in (\ref{tope}). Using 
(\ref{pg}) and
equation (\ref{app:4-int}) we obtain \begin{equation}\label{cul}
P_{g}\phi(g_1,\dots, g_4)=\sum \limits_{N_1\dots N_4} \phi{\SC 
(N_1 \dots
N_4)}^{\alpha_1 \dots \alpha_4}_{\beta_1\dots\beta_4} \  
D^{(N_1)\gamma_1}_{\alpha_1}(g_1) \dots 
D^{(N_4)\gamma_4}_{\alpha_4}(g_4)
\sum \limits_\Lambda 
C(\Lambda)_{\gamma_1\dots 
\gamma_4}C(\Lambda)^{\beta_1\dots\beta_4},
\end{equation}
See the appendix for notation and definitions.  In order to have a
more compact notation, we have dropped the indices $N_1 \dots N_4$ from
the expression for the intertwiners and we have kept only the color
$\Lambda$ that labels them (namely we use $C^{N_1 \dots N_4 \
\Lambda}_{\gamma_1\dots \gamma_4}=C(\Lambda)_{\gamma_1\dots
\gamma_4}$).  We use Einstein summation convention over repeated
indices.  There is no difference between upper or lower indices and we
write them in a way or the other for notational convenience.  We can
simplify the previous expansion by defining the new field components
\begin{equation}
\Phi_{N_1 \dots N_4, \Lambda}^{\alpha_1 \dots \alpha_4} \equiv
{\phi{\SC (N_1 \dots N_4)}^{\alpha_1 \dots
\alpha_4}_{\beta_1\dots\beta_4} C(\Lambda)^{\beta_1\dots\beta_4} \over
{(\Delta_{N_1}\dots \Delta_{N_4}})^{3 \over 2}}, 
\end{equation} 
where $\Delta_N$ denotes the dimension of the irreducible
representation of order $N$, and the factor ${(\Delta_{N_1}\dots
\Delta_{N_4}})^{3 \over 2}$ has been chosen to simplify the expression
of the interaction vertex computed below.  This choice of field yields
the mode expansion:
\begin{equation} 
    P_{g}\phi(g_1,\dots, g_4)=\sum
\limits_{N_1\dots N_4,\Lambda} (\Delta_{N_1}\dots\Delta_{N_4})^ 
{3 \over 2}\ 
\Phi_{N_1 \dots N_4,\Lambda}^{\alpha_1 \dots \alpha_4}  
\ D^{(N_1)\gamma_1}_{\alpha_1}(g_1) \dots 
D^{(N_4)\gamma_4}_{\alpha_4}(g_4)
\ C(\Lambda)_{\gamma_1\dots \gamma_4}. 
\end{equation} 

Using (\ref{app:2-int}) and the orthonormality of the intertwiners, the
kinetic term in (\ref{tope}) becomes 
\begin{equation} 
    \sK=\sum
\limits_{N_1\dots N_4,\Lambda} \Phi_{N_1\dots N_4,\Lambda}^{\alpha_1
\dots \alpha_4} \Phi_{N_1\dots N_4,\Lambda}^{\mu_1 \dots \mu_4}
(\Delta_{N_1}\dots \Delta_{N_4})^2 \ \delta_{\alpha_1\mu_1}\dots
\delta_{\alpha_4\mu_4}. 
\end{equation} 
We can directly read the propagator of the theory from this expression 
\begin{equation}
    \label{pph}
\sP_{\alpha_1\mu_1 \dots 
\alpha_4\mu_4}={\delta_{\alpha_1\mu_1}\dots
\delta_{\alpha_4\mu_4}\over (\Delta_{N_1}\dots \Delta_{N_4})^2}. 
\end{equation}

In order to write the potential term we need to express
$P_{g}P_{h}P_{g}\phi$ in terms of irreducible representations (see
(\ref{tope})).  Starting with (\ref{cul}) and using equations
(\ref{app:4-int}) and (\ref{proy}) we obtain 
\begin{eqnarray}
P_{g}P_{h}P_{g}\phi&=&\sum \limits_{N_1\dots N_4,\Lambda}
{{(\Delta_{N_1}\dots\Delta_{N_4})^{3 \over 2}}}\ \Phi_{N_1 \dots
N_4,\Lambda}^{\alpha_1 \dots \alpha_4}\ 
D^{(N_1)\gamma_1}_{\alpha_1}(g_1) \dots
D^{(N_4)\gamma_4}_{\alpha_4}(g_4) \\ &&\ \ \ \ \ \ \ \ \ \ \ \nonumber
\sum \limits_N C(N)_{\gamma_1\dots \gamma_4}
C(N)^{\beta_1\dots\beta_4} w_{\beta_1} \dots w_{\beta_4} w^{\mu_1}
\dots w^{\mu_4} C(\Lambda)_{\mu_1 \dots \mu_4}.  \end{eqnarray}
Applying equation (\ref{m}) and (\ref{bbcc}) we obtain
\begin{eqnarray}
P_{g}P_{h}P_{g}\phi&=&\sum \limits_{N_1\dots N_4,\Lambda}
 \sqrt{\Delta_{N_1}\dots \Delta_{N_4}}\ \ \Phi_{N_1 \dots
N_4,\Lambda}^{\alpha_1 \dots \alpha_4}   
D^{(N_1)\gamma_1}_{\alpha_1}(g_1)
\dots D^{(N_4)\gamma_4}_{\alpha_4}(g_4)\ \ B_{\gamma_1\dots 
\gamma_4}, 
\end{eqnarray} where the sum is now over simple representations 
only, and
$B_{\gamma_1\dots \gamma_4}$ denotes the Barrett-Crane 
intertwiner. Using the
previous equation the potential term in (\ref{tope}) becomes
\begin{eqnarray}\label{vph} \frac{1}{5!} \sum_{N_1\ldots N_{10}} 
\sum_{\Lambda_1\ldots\Lambda_5} 
\Phi_{} {}_{ N_1 N_2 N_3 N_4, 
\Lambda_1}^{\alpha_1\alpha_2\alpha_3\alpha_4} 
\Phi_{} {}_{ N_4 N_5 N_6 N_7, 
\Lambda_2}^{\alpha_4\alpha_5\alpha_6\alpha_7} 
\Phi_{} {}_{ N_7 N_3 N_8 N_9, 
\Lambda_3}^{\alpha_7\alpha_3\alpha_8\alpha_9} 
\Phi_{} {}_{ 
N_9N_6N_2N_{10},\Lambda_4}^{\alpha_9\alpha_6\alpha_2\alpha_{10}}
   \Phi_{} {}_{N_{10}N_8
N_5N_1,\Lambda_5}^{\alpha_{10}\alpha_8\alpha_5\alpha_1} ~
{{\sB}_{N_1,\ldots,N_{10}}},\end{eqnarray} where 
${\sB}_{N_1,\ldots,N_{10}}$
corresponds to a $15j$-symbol constructed with Barrett-Crane 
intertwiners
which corresponds to the Barrett-Crane vertex 
amplitude\cite{BarrettCrane}.
Explicitly, \begin{equation} \label{B2}
{\sB}_{N_1,\ldots,N_{10}} := {B}^{N_1 N_2 N_3 
N_4}_{\alpha_1\alpha_2\alpha_3\alpha_4}\ {B}^{N_4 N_5 N_6 
N_7}_{\alpha_4\alpha_5\alpha_6\alpha_7}\ {B}^{N_7 N_3 N_8 
N_9}_{\alpha_7\alpha_3\alpha_8\alpha_9}\ {B}^{N_9 N_6 N_2 
N_{10}}_{\alpha_9\alpha_6\alpha_2\alpha_{10}}\ {B}^{N_{10} 
N_8 N_5 N_1}_{\alpha_{10}\alpha_8\alpha_5\alpha_1}.
\end{equation}
Thus, the potential part of the action in the new model is given by
(\ref{vph}) as in the Barrett-Crane model.  Notice, however, that there is an
extra sum over $\Lambda$ in (\ref{vph}), absent in Barrett-Crane.  The
propagator (\ref{pph}) of the theory in momentum space is rescaled
with respect to the Barrett-Crane propagator ($ \sP_{\alpha_1 \mu_1
\dots \alpha_4 \mu_4}=(\Delta_1 \dots \Delta_4)^{-2}
\sP^{(BC)}_{\alpha_1 \mu_1 \dots \alpha_4 \mu_4}$).  As a consequence
of this rescaling, and of the extra sum over $\Lambda$, there is a 
non-trivial amplitude associated to edges in the spin foam.  Each edge
contributes to the amplitude as 
\begin{equation}\label{xii}
    A_e={\Delta_{N_1, \ldots,
    N_{4}}\over \left(\Delta_{N_1}\dots\Delta_{N_4}\right)^{2}},
\end{equation}    
where $N_1$ to $N_4$ are the colors of the four faces meeting at the
given edge, and $\Delta_{N_1, \ldots, N_{4}}$ is the dimension of the
space of the interwiners between the representations $N_1, \ldots,
N_{4}$.  In conclusion, the amplitude of a Feynman diagram $J$ is 
given by 
\begin{eqnarray}
    \label{e} 
    A(J) &=&\sum_N\  \prod_f \Delta_{N_f}\ 
\prod_e A_e \prod_v \sB_{N_{1}\dots N_{10}}. 
\end{eqnarray} 
where the sum is over simple representations $N$ of $SO(4)$, and
$\sB$ denotes the Barrett-Crane vertex amplitude.

Equivalently, as every edge connects two vertices, we can absorb the
edge amplitude in the vertex amplitude and write $Z$ in the standard
form (\ref{Z}), where the vertex amplitude is
\begin{eqnarray}
\label{vertexampli}
    A_{v} = {\prod_{i}\Delta^{1/2}_{N_{i1},\ldots, N_{i4}}
    \over 
    \left(\Delta_{N_{1}}\ldots \Delta_{N_{10}}\right)^{2}} \ 
    \sB_{N_{1}\dots N_{10}} , 
\end{eqnarray} 
where $N_{1}\ldots N_{10}$ are the ten colors of the ten faces
adjacent to the vertex $v$, and $N_{i1}\ldots N_{i4}, i=1\ldots 5$ 
are the four colors of the four faces adjacent to $i$'th edge adjacent 
to the vertex $v$. 

We close this section with a comment.  Unlikely the Barrett-Crane
model, in the mode expansion of the model presented here, the field
depends on five representations (four external and one intertwiner),
which can be seen precisely as the quantum numbers of a ``first quantized"
geometry of a tetrahedron.

\subsection{Bubbles in the spin foam formulation}

Now we are ready to show the finiteness of the 5-bubble amplitude
(equation (\ref{gordita}) in the previous section) directly in
momentum space.  As we mentioned before, the integration over
$\rho_{ab}$ in (\ref{gordita}) projects the amplitude into the trivial
representations.  The value of $\sI_5$ is then given by the sum over
colors of the amplitudes corresponding to the diagram on the right of
Fig.\ \ref{15}, in which the colors of the external faces are 
fixed to zero. Four of the
colors vanish in each of the five vertices, and a typical vertex amplitude
reduces to \begin{eqnarray} \sB_{0\,0\,0\,0\,N_1 \dots N_6}=\left\{
\begin{array}{ccc} N_1 & N_2 & N_3 \\
 N_4 & N_5 & N_6 \end{array} \right\}_{6j},\end{eqnarray}
where the RHS denotes the $6J-$symbol defined in the appendix. 
Each of the internal faces is colored by
irreducible representations $N_1$ to $N_{10}$. The 
internal faces are triangles; therefore for each $\Delta_N$,
contribution of the face, there will be 3 edge contributions
($(\Delta^{-2}_N)^3$). The factor in the numerator of (\ref{xii}) reduces to
the general form $\Delta_{0,N_1,N_2,N_3}=1$, since the space of intertwiners
between 3 representations is one dimensional. Putting this together, we get
from (\ref{e})  \begin{equation} \sI_5=\sum_{N_1\dots N_{10}}\left( 
\Delta_{N_1} \dots \Delta_{N_{10}}\right) \left(  \Delta_{N_1} \dots
\Delta_{N_{10}}\right)^{-6} \prod_v  \left\{ \begin{array}{ccc} N^v_1 & N^v_2
& N^v_3 \\  N^v_4 & N^v_5 & N^v_6 \end{array} \right\}_{6j}.\end{equation} 
Using the bound for
the $6j$-symbols given in (\ref{6J}), and noticing that 
each  color appears in three of the vertices, we conclude:  
\begin{equation} \left| \sI_5 \right| \le \sum_{N_1 \dots N_{10}} \left( 
\Delta_{N_1}
\dots \Delta_{N_{10}}\right) \left( 
\Delta_{N_1}
\dots \Delta_{N_{10}}\right)^{-6}\left( \Delta_{N_1}
\dots \Delta_{N_{10}}\right)^{3/2}= \sum_{N_1 \dots N_{10}} {\left(
\Delta_{N_1} \dots \Delta_{N_{10}}\right)^{-{7 \over 2}}}< \infty . 
\end{equation}

Next, consider the case in which the external colors are not zero. 
The finiteness of these terms follows from lemma 4 in the appendix. 
The proof follows the similar steps as before, simply replacing the
bound for the $6j$-symbols with the bound on the Barrett-Crane vertex
amplitude (\ref{cota}).  More precisely, there are two kind of colorings
corresponding to external and internal faces respectively.  We denote
by $N^e$ the colors labeling the ten external surfaces in the diagram
of Fig.\ \ref{15}, while $N^i$ denotes the colors of the ten internal
faces.  Only internal colors are summed over.  External colors label the
surfaces shown on the left of Fig.\ 5. They appear in the propagator 
corresponding to the
appropriate internal edge, thus contributing with a factor $\Delta^{-2}_{N^e}$
to the amplitude. They also appear as arguments of  two vertex amplitudes.
Using (\ref{cota}) this contribution will be less or equal than $\Delta^{2}_{N^e}$. 
There will be
no face contribution for external colors (the face contribution appears when
one has a complete chain of propagators closing around a face so that all the
$\delta_{\alpha \beta}$ in (\ref{pph}) combine into $\delta_{\alpha
\alpha}=\Delta_N$). Therefore, external colors $N^e$ can appear in the bound
for the amplitude only through the dimensionality of the space of intertwiners
$\Delta_{N_1,\dots, N_4}$ in each propagator associated to the ten internal
edges. We denote the $\Delta_{N_1,\dots, N_4}$, in (\ref{xii}), as $\Delta^{ab}$ for $a,
b=1,\dots,5$ and $a\ne b$, according to Fig.\ \ref{15}.

On the other hand,
internal colors appear as face contributions ($\Delta_{N^i}$).  There is also a
contribution from the three edges of each face ($\Delta^{-6}_{N^i}$),
and the contribution of the three corresponding vertices which are less or
equal that $\Delta_{N^i}^{3}$, according to (\ref{cota}). 

From the considerations of the two previous paragraphs we obtain
\begin{equation}
\left| A^{(bc)}_5(N^e_1 \dots
N^e_{10}) \right| \le \sum_{N^i_1 \dots N^i_{10}}
{\left( \Delta_{N^i_1} \dots
\Delta_{N^i_{10}}\right)^{-{2}}}\prod_{ab}\ \ \Delta^{ab}_{N^e,N^i}
\end{equation} 

In order to construct a manifestly finite upper bound for the amplitude we need
to  obtain a bound for $\Delta_{N_1,\dots,N_4}$. To find an upper bound for
$\Delta_{N_1,\dots,N_4}$ we proceed as follows.  The $\Lambda$'s appear in
(\ref{vph}) through $\phi^{\alpha_1 \dots \alpha_4}_{N_1N_2N_3N_4,
\Lambda_1}$, where $\Lambda$ labels the elements of an orthonormal base of
intertwiners between the four representations $N_1 \dots N_4$.  Assume that
$N_1$ is less or equal than the other $N_i$.  We chose a basis in which for
example $N_1$ and $N_2$ are in the same 3-intertwiner $C^{N_1N_2\Lambda}$ in
(\ref{basis1}).  Since $N_1$, $N_2$, and $\Lambda$ have to be $SO(4)$
compatible it follows that $N_2-N_1\le \Lambda \le N_1+N_2$.  There are
$2N_1+1=\sqrt{\Delta_{N_1}}$ $\Lambda$'s who satisfy this condition.  However,
there are additional compatibility conditions on $\Lambda$, so we conclude
that the number of possible $\Lambda$'s denoted by $\Delta_{N_1,\dots, N_4}$
satisfies $\Delta_{N_1,\dots, N_4} \le \sqrt{\Delta_{N_1}}$, since $N_1$ was
the smallest $N_i$ we can also write \begin{equation}\label{ll}
\Delta_{N_1,\dots, N_4} \le \sqrt{\Delta_{N_i}}, \end{equation}  for
$i=1,2,3,4$.  The dimensionality of the space of intertwiners is independent
of the chosen basis and the previous inequality has a basis independent
meaning. We can use the colors of the ten internal faces $N^i$
to bound the ten $\Delta^{ab}$ in the previous equation.  
Finally, according to (\ref{ll}) we obtain \begin{equation}\left|
A^{(bc)}_5(N^e_1 \dots N^e_{10}) \right| \le \sum_{N^i_1 \dots N^i_{10}}
{\left( \Delta_{N^i_1} \dots \Delta_{N^i_{10}}\right)^{-{3 \over 2}}} < \infty.
\end{equation}  The same kind of
inequality can be derived for $A^{bc}_4$.

\section{Conclusion}

We have presented in this paper a spin foam model, possibly related to 
Euclidean quantum gravity, in which the fundamental amplitudes, 
divergent in the BF and Barrett-Crane models, are finite.  The model 
is obtained as a modification of the interaction term of quantum BF
theory, formulated on the group manifold.  The modification is an 
implementation, \`a la Barrett-Crane, of the constraint that reduced 
BF to general relativity.  

We have not proven finiteness of all amplitudes at all orders in the
expansion in 2-complexes, but we suspect that by using the methods
introduced here the proof could be possible.  For each particular
amplitude, it is easy to construct a finite bound of the amplitude,
using the inequalities given in the paper.  The difficulty consists of
finding a general way of bounding the degeneracy of the vertex
amplitude given by the dimensionality of the space of intertwiners
associated to the 5 edges converging at each vertex.  In addition, the
inequalities given in the appendix might be used to study the
convergence of the full Feynman series, or sum over 2-complex.

It would be crucial to get a better understanding of the
classical limit of the model.  In particular, it would be
interesting to know whether the amplitude of a spin foam
approaches the exponential of the Einstein-Hilbert action of
the four-geometry that the spin foam approximate.   

More in general, we think that the techniques and methods introduced
here could be useful for analyzing divergences in general spin foam
models. 

\section{Acknowledgments} 
A.P. thanks FUNDACION YPF (Argentina) for its support.   This
work was partially supported by NSF Grant  PHY-9900791.

\begin{appendix}   

\section{Irreducible representations of $SO(4)$}

In this appendix we review some properties and definitions in 
the theory of irreducible representations of ${\rm SO}(4)$ (on this see
\cite{Vilenkin:1993b}), and  we state and
prove some inequalities used in the paper to show the
finiteness of amplitudes in previous sections. We follow the
conventions of \cite{dfkr}. 

Given $g \in SO(4)$  we denote by
$D^{(\Lambda)}_{\alpha\beta}(g)$ the representation matrix 
corresponding to
the irreducible representation of order $\Lambda$. Integration 
over $SO(4)$ or
the $SO(3)$ subgroup $H$ is performed with the normalized Haar 
measure of the
group and the subgroup respectively. The integration of two 
representation
matrices is given by \begin{equation} \int_{SO(4)} dg\
\overline{D^{(\Lambda)}_{\alpha\beta}(g)}  
D^{(\Lambda')}_{\alpha'\beta'}(g) =
{1\over \Delta_\Lambda} \delta^{\Lambda\Lambda'}\ 
\delta_{\alpha\alpha'}\ \delta_{\beta\beta'},
\label{app:2-int}
\end{equation}
where $\Delta(N)$ denotes the dimension of the representation.
In the case of $SO(4)$ we can choose a basis in which 
matrices are
orthogonal, and the bar can be dropped from the previous 
equation.  The 
integral of the product of three group elements is 
\begin{equation}
\label{app:3-int}  \int\limits_{SO(4)} dg \ 
D^{(N_1)}_{\alpha_1\beta_1}(g) 
 D^{(N_2)}_{\alpha_2\beta_2}(g) 
 {D^{(\Lambda)}_{\alpha\beta}(g)} = C^{N_1 
 N_2\,\Lambda}_{\alpha_1\alpha_2\alpha}\ \ {C}^{N_1 
 N_2\,\Lambda}_{\beta_1\beta_2\beta}.
\end{equation}
Here $C^{N_1 N_2 \Lambda}_{\gamma_1\gamma_2\gamma}$ are 
normalized  intertwiners (Wigner 3-j symbols) between three 
representations of ${\rm SO}(4)$; that is
$C^{N_1  N_2\,\Lambda}_{\alpha_1\alpha_2\alpha}\ \ {C}^{N_1 
N_2\,\Lambda}_{\alpha_1\alpha_2\alpha}=1$.  The intertwiner 
from the tensor product of two representations $N_1,N_2$ to 
a representation $\Lambda$, if it exists is unique.

The intertwiners between four representations are of great 
importance
in our calculation. However, in general they are not unique and
rather form a vector space. An orthonormal base can be defined 
as follows:
\begin{equation}
C^{N_1\ldots N_4\,\Lambda}_{\gamma_1\ldots \gamma_4}= \ 
\sqrt{{\rm dim}_\Lambda} \ C^{N_1 N_2 
\Lambda}_{\gamma_1\gamma_2\gamma}\ {C}^{ N_3 N_4 
\Lambda}_{\gamma_3 \gamma_4 \gamma}.
\label{basis1}
\end{equation}
With these definitions equation (\ref{app:3-int})
generalizes to the case of the integration of four 
representation matrices to
\begin{equation} \label{app:4-int}
 \int\limits_{SO(4)} dg \ D^{(N_1)}_{\alpha_1\beta_1}(g) 
 \ldots D^{(N_4)}_{\alpha_4\beta_4}(g) = \sum_\Lambda \ \ 
 C^{N_1\ldots N_4\,\Lambda}_{\alpha_1\ldots \alpha_4}\ \ 
 {C}^{N_1\ldots N_4\,\Lambda}_{\beta_1\ldots \beta_4}.
\end{equation}
Another important equation corresponds to the integration of one 
representation matrix over a
sub-group $SO(3) \subset SO(4)$, namely
\begin{equation} 
 \int\limits_{H=SO(3)} dh \ D^{(N)}_{\alpha \beta}(h) = 
w^{(N)}_{\alpha} w^{(N)}_{\beta},
\label{proy}
\end{equation}
where $w^{(N)}_{\alpha}$ represents the unit vector in the 
irreducible
representation of order $N$ left invariant by the action of the 
subgroup $H$ ($w^{(N)}_{\alpha}$ is non vanishing only if $N$ is 
simple).
Equation (\ref{proy}) defines the projector into that
one-dimensional vector space. 

As mentioned, invariant vectors exist only in 
simple 
representations. As a consequence the projection of the 
intertwiner 
${C}^{N_1\ldots N_4\,N}_{\gamma_1\ldots\gamma_4} 
w_{\gamma_1}\ldots w_{\gamma_4}$
vanishes unless all the $N_i$ and $N$ are simple. In this case 
its value is given by
\begin{equation}\label{m}
{C}^{N_1\ldots N_4\,N}_{\gamma_1\ldots\gamma_4} 
w_{\gamma_1}\ldots w_{\gamma_4} = {1\over \sqrt{\Delta (N_1) 
\ldots \Delta(N_4)}},
\end{equation}
Finally we give
the definition of the Barrett-Crane intertwiner:
\begin{equation}\label{bbcc}
B^{N_1,N_2,N_3,N_4}_{\gamma_1\dots \gamma_4} \equiv \sum_N 
{C}^{N_1\ldots
N_4\,N}_{\gamma_1\ldots\gamma_4}. \end{equation} 
The previous is the un-normalized Barrett-Crane intertwiner as 
originally
defined in \cite{BarrettCrane} which is shown to be unique up to 
scaling in
\cite{mi,bar}.

\subsection{Some lemmas}

Now we state and prove the lemma that plays an important role 
in showing
the finiteness of the 1-bubble amplitude in the modified 
Barrett-Crane model.
We define the following distribution \begin{equation}\label{ii}
\sD(g_{1},g_{2},g_{3})=\int \limits_{H^3} dh_i d\beta_i
\delta(\beta_{1}h_{1}\beta^{-1}_{1}\beta_{2}h_{2}\beta^{-1}_{2}g_{1}
\beta_{3}h_{3}\beta^{-1}_{3}\beta_{4}h_{4}\beta^{-1}_{4}g_{2}
\beta_{5}h_{5}\beta^{-1}_{5}\beta_{6}h_{6}\beta^{-1}_{6}g_{3}) 
\end{equation}
\noindent {\bf Lemma 1:}

The distribution $\sD(g_{1},g_{2},g_{3})$ is a bounded
function over $SO(4)^3$.\\

\noindent {\bf Proof:}
 In order to prove the lemma we expand (\ref{ii}) using
\begin{equation}
\delta(g)= \sum \limits_{N} \Delta(N) {\rm Tr}(D^{(N)}(g)). 
\end{equation} We integrate over $\beta_i$ and $h_i$ and make 
use of equations
(\ref{proy}) and (\ref{app:2-int}). The mode expansion of
$\sD(g_{1},g_{2},g_{3})$ results:
\begin{equation}
\sD(g_{1},g_{2},g_{3})= \sum \limits_{Simple \,\, \Lambda} {1 \over
\Delta^5(\Lambda)} {\rm Tr}(D^{(\Lambda)}(g_1g_2g_3)). 
\end{equation}
Now
\begin{equation}
\left| \sum \limits_{Simple \,\, \Lambda}
{1 \over \Delta^5(\Lambda)} {\rm Tr}(D^{(\Lambda)}(g_1g_2g_3)) 
\right| < 
\sum \limits_{Simple \,\, \Lambda}
{1 \over \Delta^5(\Lambda)} \left| {\rm 
Tr}(D^{(\Lambda)}(g_1g_2g_3)) \right| <
\sum
\limits_{Simple \,\, \Lambda} {1 \over \Delta^4(\Lambda)} < 
\infty,
\end{equation}
where we have used that in our orthogonal representation $|{\rm
Tr}(D^{\Lambda}(g))| \le \Delta_{\Lambda}$. Therefore, $\sD(g_{1},g_{2},g_{3})$
is bounded on $SO^3(4)$ $\Box$

By means of the previous lemma we were able to find a bound to 
the 1-bubble
amplitude directly in configuration space. To show the 
finiteness of the
5-bubble amplitude in the previous section, we made use of the 
mode
expansion of the amplitude and the following lemma\footnote{We 
have not
found a reference to this lemma in the literature, although we
suspect is must be a known result due to its simplicity.}. 

\noindent {\bf Lemma 2:}

The $SO(4)$ $6j$-symbols defined in terms of normalized 
intertwiners
(normalized Wigner $3j$-symbols 
$C^{N_1N_2N_3}_{\alpha_1\alpha_2\alpha_3}$ of
(\ref{app:3-int})) satisfy the following inequality:

\begin{eqnarray}\label{6J}
\left| \left\{ \begin{array}{ccc} 
N_1 & N_2 & N_3 \\
N_4 & N_5 & N_6 
\end{array} \right\}_{6j} \right| \le 
\sqrt{\Delta_{N_1}\Delta_{N_2
}\Delta_{N_3 }\Delta_{N_4}\Delta_{N_5}\Delta_{N_6}} 
,\end{eqnarray}
where $\Delta_N$ denotes the dimension of the irreducible 
representation of
order $N$.

\noindent {\bf Proof:} The $6j$-symbol is defined in terms of 
$C^{N_1N_2N_3}_{\alpha_1\alpha_2\alpha_3}$ 
as

\begin{eqnarray}
\left\{ \begin{array}{ccc} 
N_1 & N_2 & N_3 \\
N_4 & N_5 & N_6
\end{array} \right\}_{6j} \equiv 
C^{N_1N_2N_3}_{\alpha_1\alpha_2\alpha_3}
C^{N_1N_4N_6}_{\alpha_1\alpha_4\alpha_6}
C^{N_2N_4N_5}_{\alpha_2\alpha_4\alpha_5}
C^{N_3N_6N_5}_{\alpha_3\alpha_6\alpha_5},\end{eqnarray} where 
summation over
repeated indices is understood. The assertion of the lemma is 
proven by means
of calculating the following integral in two different ways.
\begin{eqnarray}
\nonumber \sS = \int dg_1 dg_2 dg_3 dg_4 &&
\left( D(g_1)^{N_1}_{\alpha_1 \beta_1} D(g_1)^{N_2}_{\alpha_2
\beta_2}D(g_1)^{N_3}_{\alpha_3 \beta_3} \right)  \left( 
D(g_2)^{N_1}_{\alpha_1
\beta_1} D(g_2)^{N_4}_{\alpha_4 \beta_4}D(g_2)^{N_6}_{\alpha_6 
\beta_6}
\right) \\  
&& \ \ \ \ \ \left( D(g_3)^{N_2}_{\alpha_2 \beta_2}
D(g_3)^{N_4}_{\alpha_4 \beta_4}D(g_3)^{N_5}_{\alpha_5 \beta_5} 
\right)  \left(
D(g_4)^{N_3}_{\alpha_3 \beta_3} D(g_4)^{N_6}_{\alpha_6
\beta_6}D(g_4)^{N_5}_{\alpha_5 \beta_5} \right). \end{eqnarray}
We can rewrite the previous equation using the representation 
property
$D(g)D(f)=D(gf)$ as
\begin{eqnarray}
\nonumber \sS &=& \nonumber  \int dg_i 
{\rm Tr}\left[D^{\SC N_1}(g_1g^{-1}_2)\right]
{\rm Tr}\left[D^{\SC N_2}(g_1g^{-1}_3)\right]
{\rm Tr}\left[D^{\SC N_3}(g_1g^{-1}_4)\right]\\
&& \ \ \ \ \ \ \ \ \ \ \ \ \ \ \ \ \ \ \ \ \ \ \ \ \ \  \ \ \ \ 
\ {\rm Tr}\left[D^{\SC
N_4}(g_2g^{-1}_3)\right] {\rm Tr}\left[D^{\SC 
N_5}(g_3g^{-1}_4)\right]
{\rm Tr}\left[D^{\SC N_6}(g_2g^{-1}_4)\right]. \end{eqnarray} 
The fact that we are
using orthogonal irreducible representations of $SO(4)$ implies 
that
$\left|{\rm Tr}[D^N(g)] \right| \le \Delta_N$. Combining this 
bound for the
trace with the normalization of the $SO(4)$ Haar measure in the 
previous
equation we conclude that
\begin{equation}
|\sS| \le {\Delta_{N_1}\Delta_{N_2 }\Delta_{N_3
}\Delta_{N_4}\Delta_{N_5}\Delta_{N_6}}. \end{equation} 
On the other hand, using equation (\ref{app:3-int}), and the 
definition of the
$6j$-symbol above we obtain \begin{eqnarray}
\sS=
\left\{ \begin{array}{ccc}
N_1 & N_2 & N_3 \\
N_4 & N_5 & N_6 
\end{array} \right\}_{6j}^2,\end{eqnarray}
which concludes the proof ${\Box}$

\noindent {\bf Lemma 3:}

The $S0(4)$ $15j$-symbols defined in terms of normalized 
intertwiners
($C^{N_1N_2N_3N_4,\Lambda}_{\alpha_1\alpha_2\alpha_3\alpha_4}$ of
(\ref{app:4-int})) satisfy the following inequality:
\begin{eqnarray}
\sum_{\Lambda_1 \dots \Lambda_{10}} \left\{ \begin{array}{ccc} 
&&\!\! N_1 \ N_2 \ N_3 \ N_4 \ N_5\\
&&\!\!N_6 \ N_7 \ N_8 \ N_9 \ N_{10}\\
&&\!\!\Lambda_1 \ \Lambda_2 \ \Lambda_3 \ \Lambda_4 \
\Lambda_5  \end{array} \right\}^2_{15j} \le 
{\Delta_{N_1}\Delta_{N_2
}\Delta_{N_3 }\Delta_{N_4}\Delta_{N_5}\Delta_{N_6} \Delta_{N_7
}\Delta_{N_8}\Delta_{N_9}\Delta_{N_{10}}} ,\end{eqnarray}
where
\begin{eqnarray}
\left\{ \begin{array}{ccc} &&\!\! N_1 \ N_2 \ N_3 \ N_4 \ N_5\\
&&\!\!N_6 \ N_7 \ N_8 \ N_9 \ N_{10}\\
&&\!\!\Lambda_1 \ \Lambda_2 \ \Lambda_3 \ \Lambda_4 \
\Lambda_5 \end{array} \right\}_{15j} := 
{C}^{\SC N_1 N_2 N_3 
N_4, \Lambda_1}_{\alpha_1\alpha_2\alpha_3\alpha_4}\,{C}^{\SC N_4 
N_5 N_6 
N_7,\Lambda_2}_{\alpha_4\alpha_5\alpha_6\alpha_7}\, {C}^{\SC N_7 
N_3 N_8 
N_9,\Lambda_3}_{\alpha_7\alpha_3\alpha_8\alpha_9}\, {C}^{\SC N_9 
N_6 N_2 
N_{10},\Lambda_4}_{\alpha_9\alpha_6\alpha_2\alpha_{10}}\, 
{C}^{\SC N_{10} 
N_8 N_5 
N_1,\Lambda_5}_{\alpha_{10}\alpha_8\alpha_5\alpha_1}.\end{eqnarray}

This lemma generalizes the previous one; however for simplicity 
we have proven
explicitly the first. The proof of the current lemma follows the 
analogous
path as the previous one with some additional indices. One 
starts by defining
an integral analogous to $\sS$ in lemma(2); namely,
\begin{eqnarray}
\label{choclo}\nonumber \sS &=& \int dg_1 dg_2 dg_3 dg_4 dg_5 
 D(g_1)^{N_1}_{\alpha_1 \beta_1} D(g_1)^{N_2}_{\alpha_2
\beta_2}D(g_1)^{N_3}_{\alpha_3 \beta_3} D(g_1)^{N_4}_{\alpha_4
\beta_4}   D(g_2)^{N_4}_{\alpha_4 \beta_4} D(g_2)^{N_5}_{\alpha_5
\beta_5}\\
&& \nonumber D(g_2)^{N_6}_{\alpha_6 \beta_6} 
D(g_2)^{N_7}_{\alpha_7
\beta_7}D(g_3)^{N_7}_{\alpha_7
\beta_7} D(g_3)^{N_3}_{\alpha_3 \beta_3}
D(g_3)^{N_8}_{\alpha_8 \beta_8}D(g_3)^{N_9}_{\alpha_9 \beta_9}  
D(g_4)^{N_9}_{\alpha_9 \beta_9}
D(g_4)^{N_6}_{\alpha_6 \beta_6}\\ 
&& 
D(g_4)^{N_2}_{\alpha_2 \beta_2}
D(g_4)^{N_{10}}_{\alpha_{10} \beta_{10}}  
D(g_5)^{N_{10}}_{\alpha_{10} \beta_{10}}
D(g_5)^{N_8}_{\alpha_8 \beta_8} 
D(g_5)^{N_5}_{\alpha_5 \beta_5}
D(g_5)^{N_1}_{\alpha_1 \beta_1}. 
\end{eqnarray} where now one has an
integration over five group variables of the appropriate product 
of twenty
representation matrices. One can relate the value of the 
integral to the sum
of $15j$-symbols squared over the $\Lambda$'s by means of 
(\ref{app:4-int}).
Finally, one finds a bound to the integral using the fundamental 
inequality
$\left| {\rm Tr}[D^N(g)] \right| \le \Delta_N$.  

Now we state two corollaries of the previous lemma. First, we 
have
that \begin{eqnarray}
\left| \left\{ \begin{array}{ccc} 
&&\!\! N_1 \ N_2 \ N_3 \ N_4 \ N_5\\
&&\!\!N_6 \ N_7 \ N_8 \ N_9 \ N_{10}\\
&&\!\!\Lambda_1 \ \Lambda_2 \ \Lambda_3 \ \Lambda_4 \
\Lambda_5  \end{array} \right\}_{15j} \right| \le 
\sqrt{\Delta_{N_1}\Delta_{N_2
}\Delta_{N_3 }\Delta_{N_4}\Delta_{N_5}\Delta_{N_6} \Delta_{N_7
}\Delta_{N_8}\Delta_{N_9}\Delta_{N_{10}}},\end{eqnarray} and 
second,
from the definition of the Barrett-Crane vertex we conclude that
\begin{equation}
\left|{\sB}_{N_1,\ldots,N_{10}} \right| \le \left( 
\sum_{\Lambda_1\dots
\Lambda_5}1 \right)\sqrt{\Delta_{N_1}\Delta_{N_2 }\Delta_{N_3
}\Delta_{N_4}\Delta_{N_5}\Delta_{N_6} \Delta_{N_7
}\Delta_{N_8}\Delta_{N_9}\Delta_{N_{10}}}. \end{equation}
An important consequence of (\ref{choclo}) is that it can also 
be used to
directly find a bound for the Barrett-Crane vertex. Take the
integral defined in (\ref{choclo}) in which the $\beta_i$ are 
not contracted
between them self as in (\ref{choclo}) but rather contracted to 
the appropriate
set of normalized invariant vectors $w^{\beta_i}$ in each 
representation. Using
(\ref{app:4-int}) and (\ref{m}) we obtain, on the one hand
\begin{equation}
{{\sB}_{N_1,\ldots,N_{10}} \over \left( \Delta_{N_1}\Delta_{N_2
}\Delta_{N_3 }\Delta_{N_4}\Delta_{N_5}\Delta_{N_6} \Delta_{N_7
}\Delta_{N_8}\Delta_{N_9}\Delta_{N_{10}}\right)}
.\end{equation}
However, on the other hand, the absolute value of integrand 
contracted with all
the $w$'s is less or equal than one since it can be written as a 
product of
terms  of the form $|w^{\mu} D(g)_{\mu\nu}w^{\mu}| \le 1$. 
Therefore, we have
proven the following lemma:\\

\noindent {\bf Lemma 4:}

The Barrett-Crane vertex amplitude satisfies the following 
inequality:
\begin{equation}\label{cota}
\left|{\sB}_{N_1,\ldots,N_{10}} \right| \le   
{\Delta_{N_1}\Delta_{N_2
}\Delta_{N_3 }\Delta_{N_4}\Delta_{N_5}\Delta_{N_6} \Delta_{N_7
}\Delta_{N_8}\Delta_{N_9}\Delta_{N_{10}}}. \end{equation}

\end{appendix}

\end{document}